\documentclass[%
 reprint,
 amsmath,amssymb,
 aps,
nofootinbib]{revtex4-2}
\usepackage{float}
\usepackage{amsmath,latexsym,amssymb,amsfonts}
\usepackage{dcolumn}
\usepackage{appendix}
\usepackage{mathtools}
\usepackage{graphicx}

\usepackage{parallel,enumitem}
\usepackage[usenames,dvipsnames]{color}

\usepackage{dcolumn}
\usepackage{bm}
\usepackage{appendix}
\usepackage{hyperref}
\usepackage{mathrsfs}
\hypersetup{
    colorlinks=true,
    linkcolor=blue,
    filecolor=magenta,      
    urlcolor=cyan,
    pdftitle={Overleaf Example},
    pdfpagemode=FullScreen,
    }
\begin{document}	
\title{\textbf{On the Generalization of the Kruskal-Szekeres Coordinates: A Global Conformal Charting of the Reissner–Nordström Spacetime}}
\author{Fawzi Aly}
\email{mabbasal[AT]buffalo.edu}
\author{Dejan Stojkovic}
\email{ds77[AT]buffalo.edu}
\affiliation{HEPCOS, Physics Department, SUNY at Buffalo, Buffalo, New York, USA}

\begin{abstract}
The Kruskal-Szekeres coordinate construction for the Schwarzschild spacetime could be interpreted simply as a squeezing of the $t$-line into a single point, at the event horizon $r=2M$. Starting from this perspective, we extend the Kruskal charting to spacetimes with two horizons, in particular the Reissner-Nordström manifold, $\mathcal{M}_{RN}$. We develop a new method to construct Kruskal-like coordinates through casting the metric in new null coordinates, and find two algebraically  distinct ways to chart $\mathcal{M}_{RN}$. We pedagogically illustrate our method by crafting two compact, conformal, and global coordinate systems labeled $\mathcal{GK_{I}}$ and $\mathcal{GK_{II}}$ as an example for each class respectively, and plot the corresponding Penrose diagrams. In both coordinates, the metric differentiability can be promoted to $C^\infty$ in a straightforward way. Finally, the conformal metric factor can be written explicitly in terms of the $t$ and $r$ functions for both types of charts. We also argued that the chart recently reported in \cite{soltani_2023} could be viewed as a type-II chart. 
\end{abstract}
\maketitle

\section{Introudction}
Reissner–Nordström (RN) spacetime is a unique, static, spherically symmetric, and asymptotically flat solution to the coupled set of Maxwell equations and Einstein field equations. It describes spacetime with the mass $M$, measured in the asymptotic region, and a static spherical electric field sourced by the charge $Q$ in the background, with the corresponding nonzero stress energy tensor. Spherical-like coordinates, $(t,r,\theta,\phi)$, known as the Reissner–Nordström coordinates are the natural coordinates to represent the metric tensor $g_{\mu \nu}$ \cite{RN_metric_2016_Jonatan_Nordebo,carroll_2003,griffiths_podolský_2012,matthias_blau,chandrasekhar_1983}. This chart could be assigned to an asymptotic observer, say Bob, at $r \rightarrow \infty$ equipped with a clock measuring the coordinate $t$. The RN metric in natural units  $(c=G=1)$ can be written as
\begin{equation}
\begin{aligned}
\mathrm{d} S_{RN}^2&=\frac{-(r-r_{+})(r-r_{-})}{r^2} \mathrm{d} t^2+\frac{r^2\mathrm{~d} r^2}{(r-r_{+})(r-r_{-})} \\
&+r^2\left(\mathrm{~d} \theta^2+\sin ^2 \theta \mathrm{d} \phi^2\right).\\
\end{aligned}
\end{equation}
 This coordinate system is ill-defined at two null hypersurfaces. Similar to the Schwarzschild spacetime, the coordinate singularity, at which $g_{tt}=0$; locates the Killing horizons of the spacetime related to the Killing vector $\partial_t$.
 \begin{equation}
\begin{gathered}
g_{tt}\left(r_{\pm}\right)=0, \\
r_{\pm}=M \pm \sqrt{M^2-Q^2}.
\end{gathered}
\end{equation}

For the non-extremal case, $M>Q$, the Reissner–Nordström black hole has an inner $r_{-}$ and outer $r_{+}$ horizon, which makes its interior somewhat similar to the interior of the Kerr Spacetime \cite{Matt_Visser,teukolsky_2015}. Further, in these coordinates the region $E_{-}=\{r|0<r<r_{-}\}$ the metric will have the same signature as in the region $E_{+}=\{r\ |\ r_{+}<r<\infty\}$. Consequently, the physical point-like singularity at $r=0$ is timelike in nature, in disagreement with the Schwardchild spacelike singularity. The metric is dynamical in the trapped and anti-trapped regions $E=\{r\ |\ r_{-}< r <r_{+}\}$ (if we consider the maximal analytical extension of the RN manifold), since the $r$ coordinate becomes timelike due to the flip of the metric signature in these coordinates \cite{matthias_blau}. 
\vspace{1mm}

We can easily illustrate the inconvenience of this chart in proximity to the RN's black hole event horizon. Let us examine Bob's clock, which times his girlfriend Alice's trip, who is for some mysterious reasons, freely falling towards the outer horizon. While Alice measures a finite amount of time, $\Delta\tau$, using her own clock in her rest frame, Bob measures a significantly dilated duration of time, $\Delta t$, by timing Alice's worldline. In other words, Bob will never see Alice crossing the outer event horizon in his lifetime. Generically, timelike (spacelike) intervals suffer from infinite dilation (contraction) once measured near the RN's horizons using Bob charting devices.  Therefore, better charts are needed to describe phenomena where Bob's tools fail \cite{MTW}. 
\vspace{1mm}

Finding a new charting system to describe regions near and across the null hypersurfaces in different spacetimes is a long-standing business. Novikov coordinates \cite{Novikovphdthesis}, Lemaître coordinates \cite{Lemaître_1997}, Gullstrand–Painlevé coordinates \cite{Martel_Poisson_2001, Robertson_Noonan_1969}, Eddington–Finkelstein coordinates \cite{Finkelstein_1958}, and Kruskal–Szekeres coordinates \cite{kruskal_1960,Unruh,lemos_silva_2021} are all examples of charts developed to overcome the incompetents of the Schwarzchild coordinates near its event horizon located at $r=2M$ where $M$ is the Schwarzchild BH mass. Some of them have been generalized to Reissner–Nordström \cite{griffiths_podolský_2012} and Kerr spacetimes \cite{Campanelli_Khanna_Laguna_Pullin_Ryan_2001,Sorge_2022}. Most of them were constructed by studying timelike and nulllike geodesics behavior around the blackhole. However, we decided to follow a different and more algebraic approach to find new global charts, relying on the simplest mathematical way to interpret what define a good coordinate. Our argument is analogous to the one found in \cite{MTW}. 
\vspace{1mm}

Although astrophysical black holes are expected to be electrically neutral \cite{Bambi_2020}, even a small amount of charge on a large black hole could be important when we encounter certain phenomena such as cosmic rays \cite{zajaček_tursunov_2019}. Also, small primordial black holes that did not live long enough to get neutralized can carry a significant amount of charge. Another exception might be black holes charged under some other hidden $U(1)$ gauge group different from electromagnetism \cite{Cardoso_2016,Dai:2009hx}. This provides enough motivation to study RN black holes not only for academic interests but also from a phenomenological point of view. On the other hand, studying the causal structure of the RN black hole, which is entirely different from the one associated with the Schwardchild spacetime, is important also as it shares some generic features with other types of black holes with two horizons, e.g. the Kerr black hole which is much more relevant in astrophysical situations \cite{teukolsky_2015,hamilton_2020}. 
\vspace{1mm}

 Kl\"{o}sch and Strobl managed to provide \emph{non-conformal} global coordinates for the extreme and non-extreme RN spacetimes \cite{klösch_strobl_1996}. Moreover, most of the attempts to construct conformal global coordinates were based on patching two sets of the Kruskal–Szekeres coordinates $\mathcal{K_{\pm}}$, where each set is well-behaved on one horizon $r_{\pm}$ while it fails on the other one $r_{\mp}$. This makes the region of validity for each chart $\mathcal{E_{+}}= E_{+} \cup \{r| r_{-} < r \leq r_{+}\}$ and $\mathcal{E_{-}}= E_{-} \cup \{r| r_{-}\leq r< r_{+}\}$ respectively. Switching between the two charts was the key to covering the whole RN manifold and constructing a global Penrose diagram in \cite{carter_1966,graves_brill_1960,hamilton_2020}. Such patched Penrose diagrams, found in \cite{matthias_blau} for example, will still prove inconvenient if we want to study geodesics across the horizons \cite{Wei_Geodesics,Fazzini_Rovelli_Soltani_2023}. To overcome this obstacle, a global conformal coordinate system is required. 
\vspace{1mm}

Recently in \cite{soltani_2023}, Farshid proposed a \emph{smoothing} technique that could be used to provide a $C^2$-conformal global chart for the RN spacetime, and pointed out the possibility of generalizing the method to spherically symmetric spacetimes. The used method was reported to be a generalization of one used by Andrew Hamilton in \cite{hamilton_2020} aiming to promote the differentiability of the map. One can also find Penrose diagrams constructed using this method in \cite{Fazzini_Rovelli_Soltani_2023}.  The central idea of this work was to find coordinates that extrapolate to each of the Kruskal–Szekeres coordinates $\mathcal{K_{\pm}}$ when approaching the horizon located at $r=r_{\pm}$. In addition, smoothing was achieved through the use of the bump functions \cite{lee_2002,tu_2008}. A similar technique was used by Schindler in \cite{Schindler_2018,Schindler_2020} to overcome the limitations of the Carter method of charting two-horizons spacetimes \cite{carter_1966}. This technique was designed to provide a global chart regular for a special class of spherically symmetric spacetimes with multiple horizons defined in the work as the \emph{Strong Spherical Symmetric} spacetimes. The reader can also find a comprehensive summary of the Penrose diagram theory in the chapter one of Schindler's doctoral thesis \cite{BHevaporation}.
\vspace{1mm}

\begin{figure}\label{PD for Single Universe}
\includegraphics[width=8cm]{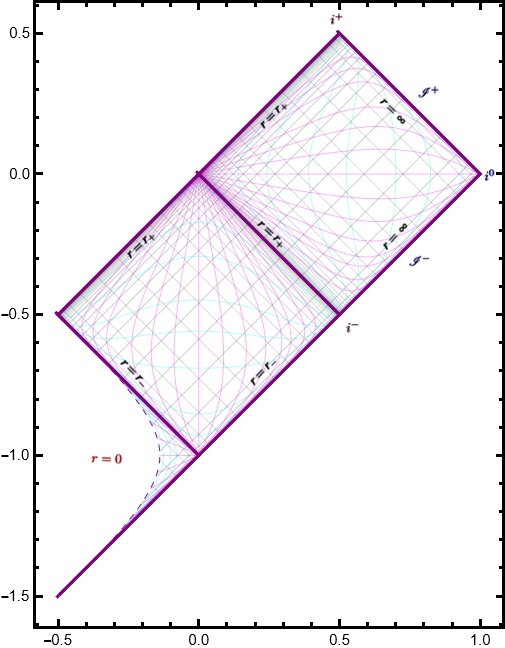}
\caption{Penrose Diagram for a single Reissner Nordstrom ($M=1$, $Q=.96$) Universe constructed using the example for Type-I coordinates given in \ref{I-Type Global Chart}. The Kruskal coordinates $(T,R)$ are plotted on the y-axis and x-axis respectively. The constant-r and constant-t curves are plotted in cyan and magenta while the null-geodesics are plotted in gray. The outer $r_{+}$ and inner $r_{-}$ horizons are described by $T=\pm R$ and $T=\pm R-1$ lines. The physical singularity at $r=0$ is plotted as a curve in dashed purple.}
\label{fig:1}
\end{figure}

\begin{figure}\label{PD for block Universe type-II}
\includegraphics[width=8cm]{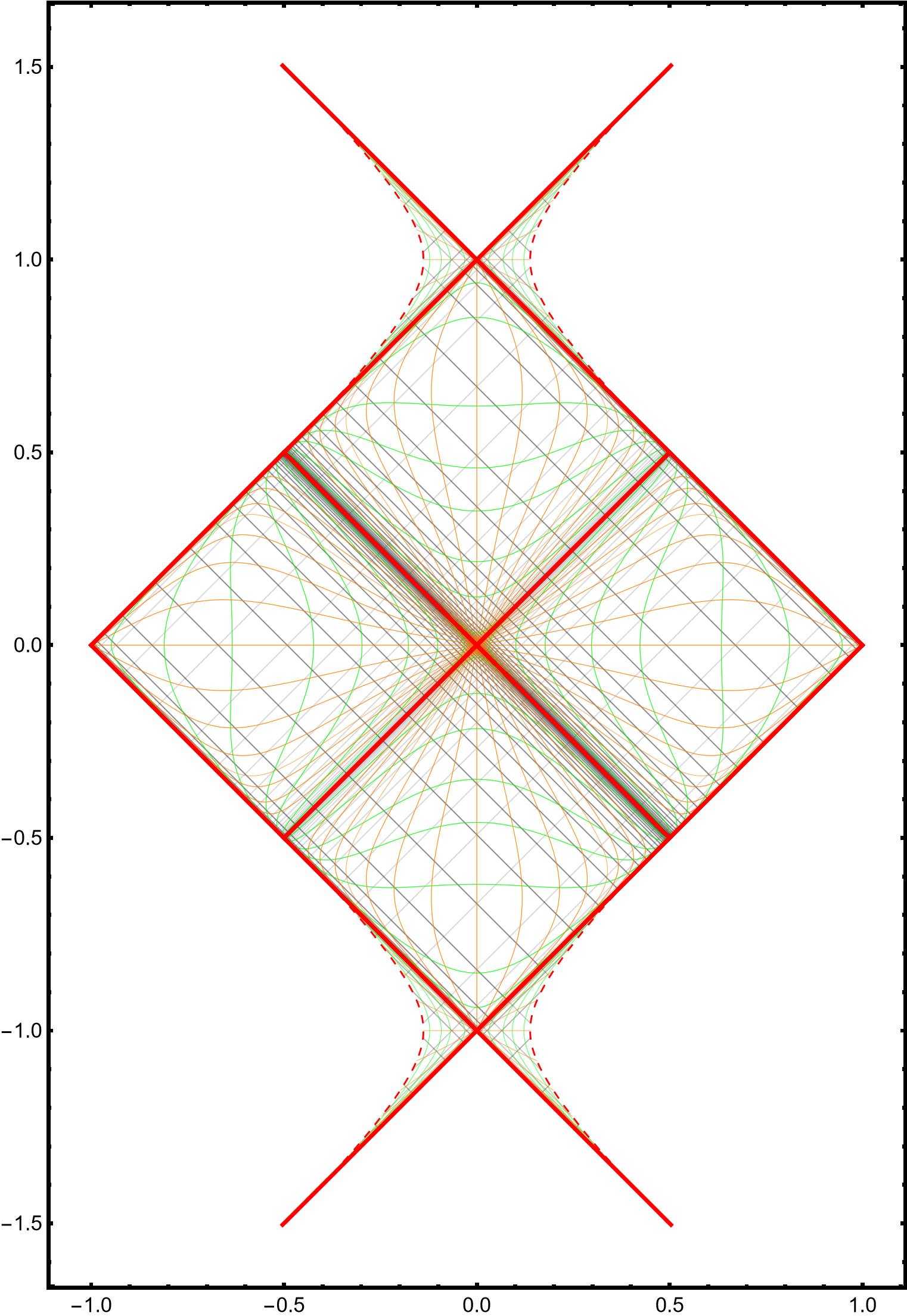}
\caption{Penrose Diagram for a block universe of Reissner Nordstrom ($M=1$, $Q=.96$) Universes using type-II coordinates found in \ref{II-Type Global Chart}. The Kruskal coordinates $(T,R)$ are plotted on the $y$-axis and $x$-axis respectively. The constant-$r$ and constant-$t$ curves are plotted in green and orange while the null-geodesics are plotted in gray. The outer $r_{+}$ and inner $r_{-}$ horizons are described by $T=\pm R$ and $T=\pm R\pm1$ lines, while the physical singularity at $r=0$ is plotted as a red curve.}
\label{2}
\end{figure}

In this work, we will define a new procedure that can produce compact, conformal, and global (CCG) charts that are valid at both the inner and outer horizons of RN spacetime, and for which the metric could be infinitely differentiable $C^\infty$. Using this procedure we will cast the possible two CCG coordinate systems for the RN spacetime into two categorizes, which we label as type-I and type-II coordinates. The reader can find a Penrose diagram of a single and Block Universe of the RN spacetime in figure [\ref{fig:1}] and figure [\ref{2}] constructed using examples of type-I and type-II CCGs respectively. Moreover, the coordinates provided in \cite{soltani_2023} could be thought of as coordinates of type-II. Our method makes no underlying assumptions about the nature of the spacetime, other than it should possess two horizons and that there exist some double null coordinates to chart it locally. Therefore, to facilitate future applications of this procedure, we will present here a detailed pedagogical approach.
\vspace{1mm}

\vspace{7mm}
The structure of this paper is as follows. In section \ref{On the Kruskal–Szekeres Coordinate}, we begin by reformulating the core idea of the Kruskal chart, and then revisit the  Kruskal charting of the Schwarzchild \ref{Kruskal Coordinate Construction: Schwardchild Spacetime Case Study} and the RN  \ref{Outer and Inner Kruskal Coordinate: Reissner Nordstrom Spacetime} spacetimes. In section \ref{Global Conformal Chart Criteria}, the main procedure for constructing generalized Kruksal charts is presented. The type-I and type-II coordinates as well as their relaxed versions for RN spacetime are given in \ref{I-Type Global Chart} and \ref{II-Type Global Chart}. Finally, we discuss the outcome of the analysis and possible future work in section \ref{Discussion}. Also the reader can find an argument why Soltani's smoothing technique could be thought of Type-II chart in Appendix \ref{Appendix A}, and an example on the relaxation step as an optional part of the procedure in Appendix \ref{Appendix B}.

\section{Preliminary}

\subsection{Kruskal–Szekeres coordinates }\label{On the Kruskal–Szekeres Coordinate}

Kruskal–Szekeres coordinates represent a maximal CCG chart for the Schwarzschild metric and has been studied extensively in the literature \cite{Unruh,lemos_silva_2021,kruskal_1960,doran_lobo_crawford_2008,KS_for_f(R)}. Their global nature is attributed to two features: (i) they can cover the null hypersurface located at radius $r=2M$ which Bob will fail to chart, and (ii) it is a maximal extension of the Schwarzchild chart representing two copies of the Schwarzschild universe. The metric written in the spherical-like coordinate known as the Schwarzschild  Spacetime $(t,r,\theta,\phi)$ \footnote{Since examining the behavior and possible problems of the spherical coordinates as $r \rightarrow \infty$ falls beyond the scope of this work, the angular dependence $(\theta, \phi)$ will be neglected from now on for simplicity.} where $t \in \mathbb{R}$, $r \in \mathbb{R}_{+}\symbol{92}\{0\}$, $\theta \in(0, \pi)$, and $\phi \in[0,2 \pi)$ takes the well-known form 
\begin{equation}
\begin{aligned}
d S_{Sch}^{2}&=\left(\frac{r-2 M}{r}\right)\left\{-d t^2+d r_{*}^{2}\right\}\\
&=\frac{1}{r(r_*)}\left(r(r_*)-2M\right)d S_{Con}^{2} ,
\end{aligned}
\end{equation}
where $d S_{Sch}^{2}$ and $d S_{Con}^{2}$ stand for the Schwarzschild and conformal metric\footnote{Conformal to 2D minkowskian manifold} respectively. Here, $r_*$ is defined\footnote{Usually, the constant of integration in defining the tortoise coordinate, $r_{*}$, is chosen to be $-2Mln(2M)$ in order to maintain dimensionless quantity inside the natural logarithm. Here, for simplicity, we omit this step. } as follows
\begin{equation}\label{tortoise schwarchild}
exp\left(r_*\right)= exp\left(r\right) \left|r - 2 M\right|^{2M}  .
\end{equation}
It is worth emphasizing that the map from $r$-coordinate to its tortoise version $r_*$ is bijective and its inverse is differentiable on each of $\mathcal{S_{+}}$ and $\mathcal{S_{-}}$ separately as defined below. This is obviously due to the modulus included in the definition of these coordinates in equation (\ref{tortoise schwarchild}). 
\vspace{1mm}

A rigorous procedure would involve solving the Einstein Field Equations in Kruskal coordinates (which is the \emph{top-down} approach as in \cite{MTW,frolov_novikov_1998}\footnote{The conformal factor in these references is written in terms of $r$, however, it is more instructive to think of $r(U,V)$ as a function of $U$ and $V$, and not as the areal coordinate $r$.}) by means of null-casting and the null gauge\footnote{freedom to redefine the null coordinates while preserving the null structure of the spacetime.}. Since the Schwarzschild coordinates cover only the regions $\mathcal{S_{-}}=\{r | 0<r<2M\}$ and $\mathcal{S_{+}}=\{r | 2M<r<\infty\}$\footnote{where only the attribution to an asymptotic observer is defined in the region $\mathcal{S_{+}}$.} of one universe of the Kruskal metric, trying to map the local chart to the global one (i.e. the \emph{bottom-up} approach) is not quite rigorous, because the map between the two charts as well as the Jacobian, Hessian, and the higher-versions of it will be singular at the event horizon \cite{Mathmatica_notebook}.
\vspace{1mm}

Nevertheless, we seek a global chart in which the metric is at least $C^2$ everywhere on the manifold in order to satisfy the coupled Field Equations which contain first and second derivatives of the metric. Thus, we can apply this bottom-up approach (as in most of the General Relativity textbooks \cite{carroll_2003,Wald}) by studying the limit at $r=2M$ and analytically continuing the metric there. Finally, the metric $g_{\mu \nu}$ must be written explicitly in the Kruskal coordinates $(T,R,\theta,\phi)$ only. In this paper, we will follow the bottom-up approach to find the generalized Kruksal coordinates which chart the whole RN spacetime. Taking the Kruskal charting of the Schwarzschild black hole as our guide, we review the traditional derivation of the Kruskal coordinates.

\subsection{Construction of the Kruskal coordinates: Schwardchild Spacetime}\label{Kruskal Coordinate Construction: Schwardchild Spacetime Case Study}

We begin by mapping the Schwarzschild coordinates to intermediate null coordinates first, in particular the retarded ($u$) and advanced ($v$) time coordinates, defined as $u=t-r_*$ and $v=t+r_*$. To handle the coordinate singularity of the former at the horizon, $r=2M$, the null freedom is used to map the latter set to another set of the null coordinates using $u \rightarrow U \equiv h(u)$ and $v \rightarrow V \equiv k(v)$. This gives
\begin{equation}
d S_{con}^{2}=-du dv=- \dfrac{d U d V}{\dfrac{dh}{du}\dfrac{dk}{dv}}\equiv- \dfrac{Q(U,V)d U d V}{r(U,V)-2M} ,
\end{equation}
where $Q(U,V)$ is at least $C^2$-function  $\mathcal{S}=\mathcal{S_{+}}\cup \mathcal{S_{-}} \cup \{r|r=2M\}$. This is achieved by employing the definition of $r_*$.  A sufficient coordinate transformation is given by

\begin{equation}
    \begin{aligned}
    \displaystyle
        &U \equiv \nu exp\left(\dfrac{-u}{4M}\right),
        &V \equiv \nu exp\left(\dfrac{v}{4M}\right),
    \end{aligned}
\end{equation}
where 
\begin{equation}
\nu=
     \begin{dcases}
        +1 & r>2M \\
        -1 & r<2M \\
    \end{dcases}
    ,
\end{equation}

The signs $\pm$ are included to achieve the maximal analytical extension of the metric. The product $UV$ is positive in the regions II and III, and negative in the regions I and IV, following the convention given in \cite{carroll_2003}. The $r$ coordinate is defined implicitly as 
\begin{equation}
 UV=exp\left(\frac{r}{2M}\right)(r-2 M).
\end{equation}
This equation can be explicitly solved for $r$ by employing the multi-valued Lambert function $W$ \cite{corless_gonnet_hare_jeffrey_knuth_1996,MG2010},
\begin{equation}
r=2M \left[W\left(\frac{UV}{-2Me}\right)+1\right].
\end{equation}
where $e$ is the Euler number. Then, the Schwarzchild metric will have the following form in the new double null coordinates
\begin{equation}\label{schwardchild_Kruskal_UV}
d S_{Sch}^2=-\frac{16 M^2 e^{-\dfrac{r(U,V)}{2 M}}}{r(U,V)} d U d V+ r^2(U,V) d \Omega^2.
\end{equation}
Finally, the Kruskal coordinates $T_{KS}$ and $R_{KS}$ are related to the new null coordinates through the following transformations
\begin{equation}\label{Kruskal_time_space_SCH}
\begin{array}{l}
U\equiv\frac{1}{2} \left(T_{KS}-R_{KS}\right), \\
V\equiv\frac{1}{2} \left(T_{KS}+R_{KS}\right).
\end{array}
\end{equation}
It is worth writing the final version of the metric in the Kruskal coordinates as
\begin{equation}\label{schwardchild_Kruskal}
\begin{aligned}
d S_{Sch}^2&= \omega(T_{KS},R_{KS})(-d T_{KS}^2 + d R_{KS}^2)\\
&+ 4M^2\left( W\left(T_{KS},R_{KS}\right)+1\right)^2 d \Omega^2,\\
\end{aligned}
\end{equation}
where
\begin{equation}
 \omega(T_{KS},R_{KS})=\frac{8 M exp\left[-W\left(T_{KS},R_{KS}\right)-1\right]}{W\left(T_{KS},R_{KS}\right)+1}.
\end{equation}

As a cross-check, one could verify that the Einstein tensor $G_{\mu \nu}$ corresponding to the Kruskal metric is zero everywhere on the Schwarzschild manifold, thus confirming that the stress-energy tensor $T_{\mu \nu}$ is identically zero (as it must be for the Schwarzschild solution). This is true despite the fact that taking the derivatives of the metric with respect to the coordinates $(T,R)$ (using implicit differentiation with respect to $(t,r)$) will be ill-defined at the event horizon. One could also verify that the maps between the Kruskal and the Schwarzschild chart are diffeomorphic in the regions $S_{+}$ and $S_{-}$ \cite{Mathmatica_notebook}.

\subsection{A simple interpretation of the Kruskal charting}

\begin{figure}[H]
\includegraphics[width=8.5cm]{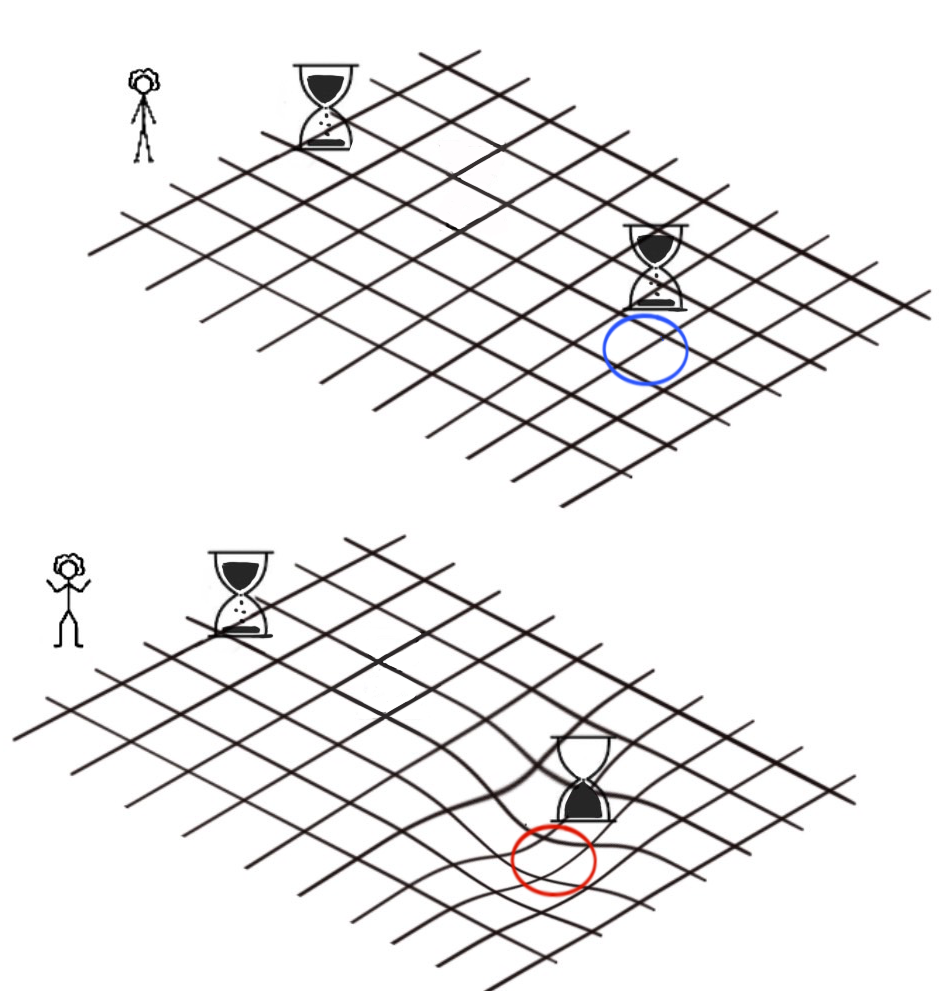}
\caption{Bob, an asymptotic observer; is trying to time out the duration of an event (or the time interval between two successive events) taking place in two different locations. In the upper picture, there is no curvature, and Bob's clock timed both events in finite and equal intervals of time. However, when Bob uses the same clock in the lower picture with a curved spacetime, the time interval keeps getting longer as the event take place near to the event horizon, until the interval becomes infinite exactly at the horizon (the red circle). Credit for this illustration to the artist Haidi Fawzi }
\label{Bob's Clock}
\end{figure}

The procedure of constructing Kruskal coordinates for Schwarzschild spacetime outlined in the previous section becomes limited when applied to spacetimes with more than one horizon. To be able to resolve this obstacle, we re-interpret the main premise of the construction. If Bob lived in a four-dimensional Minkowski spacetime, his clock would be able to properly time the events taking place there globally. However, once the spacetime is only asymptotically Minkowskian, the chart will fail near the null hypersurfaces. We can illustrate this scenario through the cartoon shown in figure [\ref{Bob's Clock}].

\begin{figure}[H]
\includegraphics[width=8.5cm]{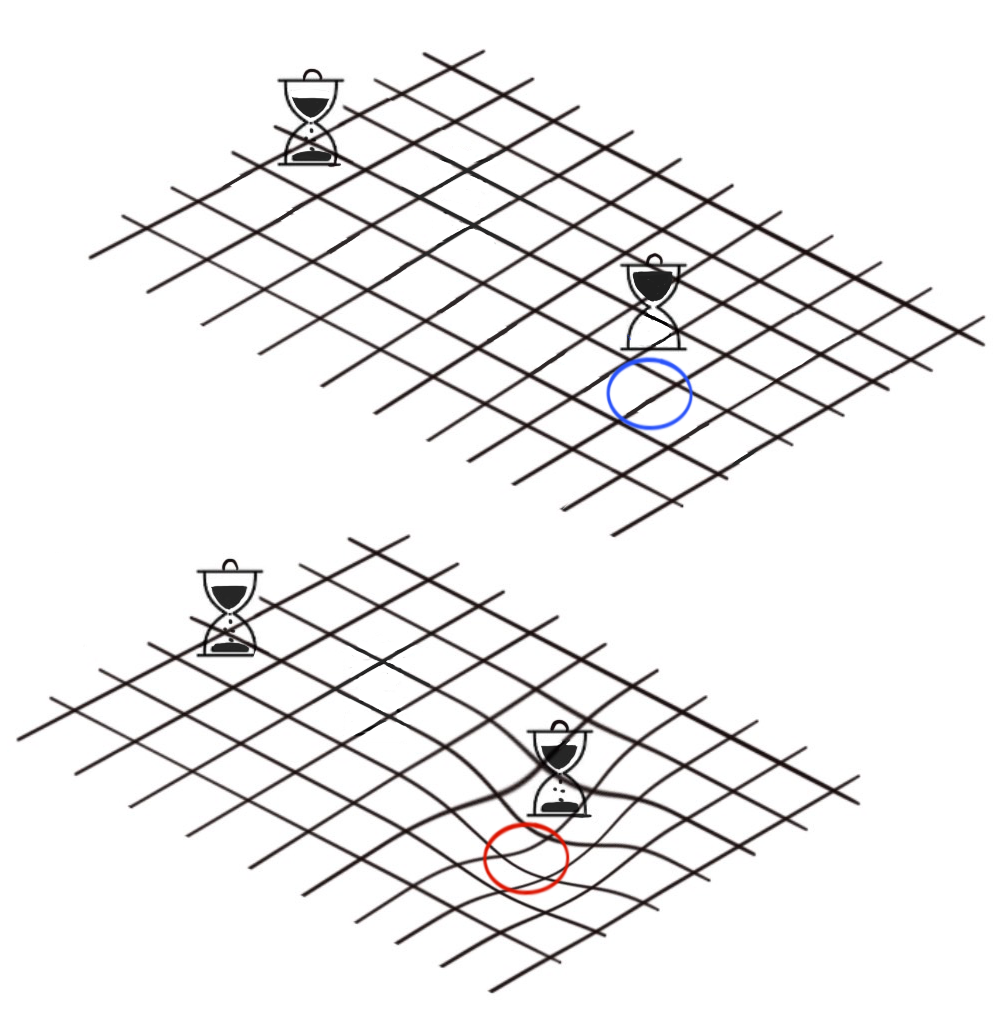}
\caption{In the upper flat spacetime, the clock is constructed in a way to squeeze the time interval to zero at the particular location where the horizon would be located in a curved spacetime (the blue circle). Consequently, if Bob used this weird clock in a flat universe, the measurements won't make any sense. However, if the same Kruskal's clock is used in the curved spacetime, it will report finite time intervals in proximity and across the horizon (the red circle). Credit for this illustration to the artist Haidi Fawzi  }
\label{Kruskal's Clock}
\end{figure}

\emph{But what if the charting method (the ruler and the clock) were ill-defined in a Minkowski spacetime at precisely these locations which correspond to the null hypersurfaces in a curved spacetime?} For example, we can define a "bad" chart $\mathcal{Z}$ in the conformal spacetime with the metric $g^{Con}_{\mu \nu}$, in which any given timing of $\Delta\tau$ of Alice's trip to the $r=2M$\footnote{In our analysis the conformal spacetime is Minkowski, so there are no horizons at the radius $r=2M$.} is mapped to $\Delta\tilde{t}\rightarrow 0$. Apparently, there is a family of these ``\emph{bad}" charts $\mathcal{Z}$ that would be well defined on the physical spacetime, with the metric $g_{\mu \nu}=\omega(x) g^{Con}_{\mu \nu}$, where $\omega(x)$ is the conformal factor. They are only conditioned to contract the time interval $\Delta \tau$ at the same rate as the dilation of time in Bob's frame. One can find an equivalent argument in \cite{MTW} that we quote here "\emph{A better coordinate system, one begins to believe, will take these two points at infinity and spread them out into a line in a new ($r_{new},t_{new}$)-plane; and will squeeze the line ($r=2M, t$ from $-\infty$ to $\infty$) into a single point in the ($r_{new},t_{new}$)-plane}". A cartoon similar to figure [\ref{Bob's Clock}] can be made for such clocks which we will name \emph{Kruskal's clock} in figure [\ref{Kruskal's Clock}]. In this step, we are not associating such a chart with any particular observer. 

As we will show here later, applying this simple argument to spacetimes with more than one horizon would be a tedious algebraic task. Mathematically, the fundamental premise of the construction is to find conformal coordinates $\mathcal{Z}$ that generate poles of the same rank as zeros of the conformal factor. Then in this bottom-up approach the zeros and poles cancel out, and the physical metric in $\mathcal{Z}$ will be CCG. In the next subsections, we will review the Kruskal charting of the RN spacetime following the notation in \cite{griffiths_podolský_2012, hamilton_2020}. 

\subsection{Outer and inner Kruskal coordinatse: Reissner-Nordstrom spacetime}\label{Outer and Inner Kruskal Coordinate: Reissner Nordstrom Spacetime}
 One example where the standard Schwarzchild-like Kruskal charting will fail in constructing a CCG is the RN spacetime. 
\begin{equation}\label{RN Metric in RN chart}
\begin{aligned}
d S_{RN}^{2}&=\frac{\left(r-r_{+}\right)\left(r-r_{-}\right)}{r^2}\left\{-d t^2+d r_{*}^2\right\}\\
&=\frac{\left(r-r_{+}\right)\left(r-r_{-}\right)}{r^2}d S_{Con}^{2},\\
\end{aligned}
\end{equation}
where $d S_{RN}^{2}$ stands for RN metric, while $(u,v)$ represents the double null coordinates constructed in the same manner as in the Schwardchild case. The RN radial tortoise coordinate $r_*$ is defined as 
\begin{equation}\label{RN tortoise coordinate}
\begin{gathered}
 exp\left(\large r_{*}\right)=exp\left( r\right)\left|r-r_{+}\right|^{\dfrac{\alpha_{+}}{2}}\left|r-r_{-}\right|^{-\dfrac{\alpha_{-}}{2}}, \\
 \alpha_{+}\equiv \frac{2 r_{+}^2}{r_{+}-r_{-}}, \\
 \alpha_{-}\equiv \frac{2r_{-}^2}{r_{+}-r_{-}},\\
\end{gathered}
\end{equation}
 where $\alpha_{-}$ and $ \alpha_{+}$ are the surfaces gravity at $r_{-}$ and $r_{+}$ respectively.
 \vspace{1mm}
 
 Similar to the Schwardchild tortoise coordinate, $r_*(r)$ is bijective and its inverse is differentiable on $E_{+}$, $E$, and $E_{-}$ separately.  However, there is a potential to solve explicitly for $r$ by employing generalized Lambert functions $\mathcal{W}$ \cite{Mezo_Baricz_2017,Mezo_Corcino_Corcino_2020,Mezo_Keady_2016,Scott_Mann_Martinez_II_2006}. Since this is a tedious task on its own, we confine our analysis to the main objective, while this step could be addressed in future work.
 \vspace{1mm}

 By examining the tortoise coordinate definition, it is obvious that a zero at $r_{\pm}$ is always coupled with a pole at $r_{\mp}$, hence it is not straightforward to factor out a product of simple poles at $r_{+}$ and $r_{-}$ in the conformal metric. Nevertheless, it remains possible to construct regular charts at one horizon that is ill-defined at the other. These coordinates are regular in the domain $\mathcal{E_{+}}$ and $\mathcal{E_{-}}$, respectively. The outer $\mathcal{K_{+}}$ and inner $\mathcal{K_{-}}$ Kruskal coordinates are simply related to the ``$-$" null-chart $(\mathcal{U}_{-}, \mathcal{V}_{-})$ and ``$+$" null-chart $(\mathcal{U}_{+}, \mathcal{V}_{+})$ following the similar but unitless definitions as those found in (\ref{Kruskal_time_space_SCH}). We will work with the following sign convention
\begin{equation}\label{Definition of U and V +}
\begin{aligned}
    \mathcal{U}_{+}&= \nu_{+} U_{+},\quad\\
    U_{+}&=exp \left(\frac{-u}{\alpha_{+}}\right),\quad\quad\\
\end{aligned}
\begin{aligned}
     \mathcal{V}_{+}&= \nu_{+} V_{+}\\
     V_{+}&=exp \left(\frac{v}{\alpha_{+}}\right),\\
\end{aligned}
\end{equation}
where 
\begin{equation}
\nu_{+}=
     \begin{dcases}
        +1 & r>r_{+} \\
        -1 & r<r_{+} \\
    \end{dcases}
,
\end{equation}
to represent the maximal analytical extension of these coordinates. Then the $t$ and $r$ coordinates are characterized by the following curves in the $(\mathcal{U}_{+},\mathcal{V}_{+})$-plane:
\begin{equation}\label{radialplus}
\begin{gathered}
\mathcal{U}_{+} \mathcal{V}_{+}=exp\left(\frac{r}{2 \alpha_{+}}\right)\left(r-r_{+}\right)\left|r-r_{-}\right|^{-\alpha},\\
\frac{\mathcal{V}_{ +}}{\mathcal{U}_{ +}}=\pm exp\left( + \dfrac{2 t}{\alpha_{ +}}\right).
\end{gathered}
\end{equation}
Similarly,
\begin{equation}\label{Definition of U and V -}
\begin{aligned}
    \mathcal{U}_{-}&= \nu_{-} U_{-},\quad\\
    U_{-}&=exp \left(\frac{u}{\alpha_{-}}\right),\quad\quad\\
\end{aligned}
\begin{aligned}
     \mathcal{V}_{-}&= \nu_{-} V_{-},\quad\\
     V_{-}&=exp \left(\frac{-v}{\alpha_{-}}\right)\\ 
\end{aligned}
\end{equation}
where \begin{equation}
\nu_{-}=
     \begin{dcases}
        +1 & r>r_{-} \\
        -1 & r<r_{-} . 
    \end{dcases}
\end{equation}
The $t$ and $r$ curves in the $(\mathcal{U}_{-},\mathcal{V}_{-})$-plane are defined as 
\begin{equation}\label{radialminus}
\begin{gathered}
\mathcal{U}_{-} \mathcal{V}_{-}=exp\left(-\frac{r}{2 \alpha_{-}}\right)\left(r-r_{-}\right)\left|r-r_{+}\right|^{-\bar{\alpha}},\\
\frac{\mathcal{V}_{ -}}{\mathcal{U}_{ -}}=\pm exp\left( - \dfrac{2 t}{\alpha_{-}}\right),
\end{gathered}
\end{equation}
Consequently, the metric in these ``+" or ``-" null-charts becomes
\begin{equation}\label{Metric in + and -}
\begin{aligned}
d S_{R N}^2&= -\alpha_{ \pm} \frac{\left(r-r_{+}\right)\left(r-r_{-}\right)}{r^2} \frac{d \mathcal{U}_{ \pm} d \mathcal{V}_{ \pm}}{\mathcal{U}_{ \pm} \mathcal{V}_{ \pm}}\\
&=-\alpha_{+} \frac{exp\left(-\dfrac{2 r}{\alpha_{+}}\right)}{r^2} \left(r-r_{-}\right)^{1+\alpha}  d \mathcal{U}_{+} d \mathcal{V}_{+}\\
&=-\alpha_{-} \frac{exp\left(\dfrac{2 r}{\alpha_{-}}\right)}{r^2} \left(r_{+}-r\right)^{1+\bar{\alpha}}d \mathcal{U}_{-} d \mathcal{V}_{-},\\
\end{aligned}
x`x`\end{equation}
where\footnote{The extreme cases ($Q=M$ and $Q>M$) of the RN metric are not considered here.}
\begin{equation}
\begin{aligned}
&\alpha \equiv \frac{\alpha_{-}}{\alpha_{+}}=\left(\frac{r_{-}}{r_{+}}\right)^2 \rightarrow 
& 0<\alpha<1,\\
&\bar{\alpha} \equiv \frac{\alpha_{+}}{\alpha_{-}}=\left(\frac{r_{+}}{r_{-}}\right)^2 \rightarrow 
& 1<\bar{\alpha}.
\end{aligned}
\end{equation}
It is easy to check that the metric in ``$+$" (``$-$") null-coordinates is regular at the outer (inner) horizon $r_{+}$($r_{-})$. However, the coordinates fail \footnote{$dS^{2}_{RN}=0$ at $r=r_{-}$ ($r_{+}$) in the ``$+$" (``$-$") coordinates according to equation (\ref{Metric in + and -}).} at the inner (outer) horizon $r_{-}$ ($r_{+}$). Moreover, the metric in the ``$+$"-null coordinates is not asymptotically flat in agreement with the Schwarzschild induced metric defined on the hypersurfaces with fixed $\theta$ and $\phi$ in equation (\ref{schwardchild_Kruskal}), where the conformal factor approaches zero as $r\rightarrow\infty$. Nevertheless, global Kruskal coordinates could be built by combining these two definitions in (\ref{Definition of U and V +}, \ref{Definition of U and V -}) together (see e.g. the work of Carter\cite{carter_1966}, Hamilton\cite{hamilton_2020} Schindler\cite{Schindler_2018}, and Farshid \cite{soltani_2023}). Although they all managed to find a regular metric across the horizon, yet the metric at maximum was only $C^2$.

\section{Global Conformal Chart Criteria}\label{Global Conformal Chart Criteria}

We start our analysis by studying the conditions needed for a valid conformal global chart. We want to map the double null coordinates $(u,v)$ to the global double null coordinates $(\tilde{u},\tilde{v})$. The most direct way to achieve that will be to use only the null gauge \cite{Wei_private,wei_2023} as follows
\begin{equation}
\begin{aligned}
&\tilde{u} \equiv h(u) \quad \rightarrow
& du= \frac{1}{\dfrac{dh}{du}} d\tilde{u}, \\
&\tilde{v} \equiv k(v) \quad \rightarrow
& dv= \frac{1}{\dfrac{dk}{dv}} d\tilde{v}. \\
\end{aligned}
\end{equation}
To construct a well-defined chart on the entire Reissner–Nordström manifold, we identify three distinct possibilities with reference to the singularity structure of the term $\dfrac{dh}{du}\dfrac{dk}{dv}$, focusing on its behavior at $r=r_-$ and $r=r_+$. The three options are:
\begin{enumerate}  
    \item  \emph{\textbf{Type-O}}: $\dfrac{dh}{du}\dfrac{dk}{dv}$ has a zero either at $r=r_-$ or $r=r_+$.  
     The regularity of the metric in the new $(\tilde{u},\tilde{v})$ coordinates would be achieved at $r=r_{-}$ or $r=r_{+}$ but not simultaneously. ``$\pm$" null coordinates are examples for this case. However, generating nontrivial coordinates out of $(U_\pm, V_\pm)$ is possible\footnote{The transformations that lead to such coordinates are expected to be more complicated as they are restricted by the requirement to leave the singularity structure invariant or to generate a decoupled zero at the other horizon.} \footnote{For example, the transformation of the form $f(U)\equiv U^p$ and $g(V) \equiv V^p$ where  $p \in \mathbb{R} $ are allowed since they clearly change the singularity structure of the term $\dfrac{dh}{du}\dfrac{dk}{dv}$.}. This condition could be formulated as follows
     \begin{equation}
\frac{d h}{d u} \frac{d k}{d v}=\left(r-r_{\pm}\right)\zeta\left(r_*, t\right)
\end{equation}
    
    \item \emph{\textbf{Type-I}}: $\dfrac{dh}{du}\dfrac{dk}{dv}$ has a product of zeros at $r=r_{-}$ and $r=r_{+}$.

    If we manage to factor out this product of zeros while keeping the associated poles decoupled, then we will have a conformal global coordinate for the RN spacetime. We will illustrate this case with an example in \ref{I-Type Global Chart}. This condition can be formulated as 
    \begin{equation}
\quad\quad\frac{d h}{d u} \frac{d k}{d v}=\left(r-r_{+}\right)\left(r-r_{-}\right) \gamma\left(r_*, t\right)
\end{equation}
    \item \emph{\textbf{Type-II}}: a sum of decoupled simple zeros at $r=r_{+}$ and $r=r_{-}$, each coupled to a pole, and possibly zeros of constrained rank at $r=r_{-}$ for former and $r=r_{+}$ for later. In principle, this mixture of poles and zeros might be easier to find compared to \emph{\textbf{Type-I}}, however, the metric is expected to be more complicated form-wise. We will illustrate this case with an example in \ref{II-Type Global Chart}. This condition can be formulated as 
      \begin{equation}
        \begin{aligned}
             \quad\quad\frac{dh}{du}\frac{dk}{dv}&=(r-r_{+}){M_{+}(r_{*},t)}+(r-r_{-})M_{-}(r_{*},t)\\
             &+\beta(r_{*},t),
        \end{aligned}
    \end{equation}
\end{enumerate}
The three differential equations listed above are sufficient to construct the desired singularity structure in each case, while the constraints are encoded within $\zeta$, $\gamma$, $M_{\pm}$ and $\beta$.

\section{Constructing CCGs for Reissner–Nordström spacetime}

\subsection{Type-I CCG Global Chart}\label{I-Type Global Chart}

\begin{figure}
\includegraphics[width=8cm]{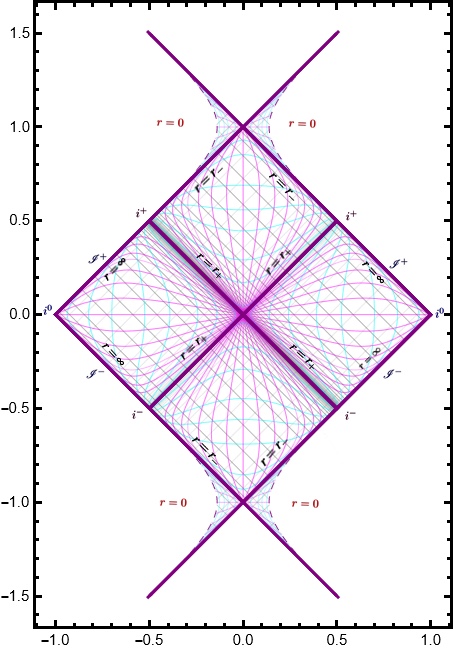}
\caption{Penrose Diagram for a block of Reissner Nordstrom ($M=1$, $Q=.96$) Universes. The Kruskal coordinates $(T,R)$ are plotted on the y-axis and x-axis respectively. The constant-r and constant-t curves are plotted in cyan and magenta while the null-geodesics are plotted in gray. The outer $r_{+}$ and inner $r_{-}$ horizons are described by $T=\pm R$ and $T=\pm R\pm1$ lines. While the physical singularity at $r=0$ is plotted as a dashed purple curve.}
\label{PD for Block Universe}
\end{figure}

As we mentioned before, just by looking at the definition of $r_*$, there is no simple way of factorizing the zeros $(r-r_{+})$ or $(r-r_{-})$ without invoking poles at $r_{-}$ or $r_{+}$. Still, we can consider combining equations (\ref{radialminus}) and (\ref{radialplus}) 
\begin{equation}
\mathcal{U}_{+} \mathcal{U}_{-} \mathcal{V}_{+} \mathcal{V}_{-}=\frac{\left(r-r_{+}\right)}{\left|r- r_{+}\right|^{\bar{\alpha}}} \frac{\left(r- r_{-}\right)}{\left|r-r_{-}\right|^{\alpha}} exp \left({\dfrac{r}{2 \alpha_{+}}}-\dfrac{r}{2 \alpha_{-}}\right)
\end{equation}
This may give us a hint for how to find $(\tilde{u},\tilde{v})$ with the desired map to fulfill the singularity structure of type-I. For example, we can start with the following definitions of $\mathcal{GK_{I}}$

\begin{equation}\label{I-Type}
\begin{aligned}
\frac{d h}{d u}=\frac{\mu_u}{U_{+}^{-1}+U_{-}^{-1} },\quad\quad\quad
\end{aligned}
\begin{aligned}
\frac{d k}{d v}=\frac{\mu_v}{V_{+}^{-1}+V_{-}^{-1} },\quad\quad\quad
\end{aligned}
\end{equation}
where can define a sign function $\mu$
\begin{equation}\label{mu def}
\mu=
     \begin{dcases}
        +1 & r>r_{+} \mid\hspace{1mm}\mid r<r_{-} \\
        -1 & r_{-}<r<r_{+}. \\
    \end{dcases}
\end{equation}
The definition we give in (\ref{I-Type}) reduces to evaluating the $I_{1}$-integration given here 
\begin{equation}\label{integation I_{1}}
I_{1}=\int \frac{1}{x^q+1} d x,
\end{equation}
where $q>1$.
This integration has an upper and lower bound, hence the sign convention we use here will locate the inner horizon $r_{-}$, outer horizon $r_{+}$ and the asymptotically flat region $r\rightarrow \infty$ according to the choice of the reference point. We choose that point to be the outer horizon $u\rightarrow-\infty$ ($v\rightarrow\infty$). Accordingly, if we stick to the $\mu=\mu_v=\mu_u$ in (\ref{mu def}), we can have a monotonic map from $u$($v$), defined in any of the regions $E_{\pm}$ and $E$; to $\tilde{u}$ ($\tilde{v}$). However, as we want to explicitly chart one of the RN universe let's choose $\mu_v=1$ while $\mu=\mu_u$. As the ODEs in (\ref{I-Type}) could be easily integrated in terms of the Hypergeometric functions ${}_2 F_{1}(a,b;c;x)$ \cite{ronveaux_2007,hypergeometric1,hypergeometric2}. 
\begin{equation}\label{definitions of type-I in HyperGeo}
\tilde{u}=\left\{\begin{array}{l}
\frac{S_{-}(u)}{S_0}-1 \quad\quad r>r_{+} \\
1-\frac{S_{-}(u)}{S_0} \quad\quad r_{+}>r>r_{-}\\
\frac{S_{-}(u)}{S_0}-2 \quad\quad r_{-}>r \\
\end{array}\right.
\end{equation}
and
\begin{equation}
\tilde{v}=\frac{S_{+}(v)}{S_0}-1
\end{equation}
while
\begin{equation}
\begin{aligned}
    S_{ \pm}(x)=&\alpha_{ \pm} e^{\frac{x}{\alpha_{ \pm}}}{ }_2 F_1\left(1, k_{ \pm}, 1+k_{ \pm},-e^{\kappa x }\right)\\
    S_{0}=&max[S_{ \pm}(x)]\\
\end{aligned}
\end{equation}
and
\begin{equation}
\begin{aligned}
    k_{ \pm}=&\frac{\alpha_{\pm}}{\alpha_{+}+\alpha_{-}}\\
    \kappa=&\frac{1}{\alpha_{+}}+\frac{1}{\alpha_{-}}\\
\end{aligned}
\end{equation}
These coordinates have a \emph{compact} domain and, hence, could be used directly to construct Penrose diagrams for the RN spacetime. The choice of signs $\mu_v$ and $\mu_u$ does not harm the continuity or differentiability of the map, still, it will cause signature flip of the metric once written in those Generalized null Kruskal Coordinate of type-I, however this is straightforwardly treatable by absorbing the signs while defining their $(T_{KS},R_{KS})$ version. The function of  $\mu_v$ and $\mu_u$  is simply to define the maximal analytical extension of RN manifold. The metric in the CCG type-I coordinates will take the following form:
\begin{equation}\label{I-Type Metric}
\begin{aligned}
&d S_{RN}^2
=-\mu_v \mu_u \frac{d \tilde{u} d \tilde{v}}{r^2}\left \{\left|r-r_{-}\right|^{\alpha+1} exp\left(-\frac{2 r}{\alpha_{+}}\right)\right.\\
&+\left.\left|r-r_{+}\right|^{\bar{\alpha}+1} exp\left(\frac{2 r}{\alpha_{-}}\right)+ 2 \cosh\left[t\left(\frac{1}{\alpha_{+}}+\frac{1}{\alpha_{-}}\right)\right]\right.\\
&\left. \times exp\left(r\left[\frac{-1}{\alpha_{+}}+\frac{1}{\alpha_{-}}\right]\right)\left|r-r_{+}\right|^{\frac{\bar{\alpha}+1}{2}} \left|r-r_{-}\right|^{\frac{\alpha+1}{2}} \right \}.\\
\end{aligned}
\end{equation}
Using these coordinates a block Penrose diagram is constructed as shown in figure [\ref{PD for Block Universe}].

. 

The metric possesses a conformal factor resembling the sum of the conformal factors of the $\mathcal{K_{\pm}}$ in addition to a new time-dependent term that vanishes on both horizons. The metric is well-behaved on both of the horizons and takes the following asymptotic behavior as $r\rightarrow r_{+}$,
\begin{equation}
d S_{RN}^2 (r\rightarrow r_{+})\rightarrow -\frac{exp\left(-\dfrac{2 r}{\alpha_{+}}\right)}{r^2} \left(r-r_{-}\right)^{1+\alpha}d \tilde{u} d \tilde{v},
\end{equation}
Similarly as $r\rightarrow r_{-}$,
\begin{equation}
d S_{RN}^2 (r\rightarrow r_{-})\rightarrow - \frac{exp\left(\dfrac{2 r}{\alpha_{-}}\right)}{r^2} \left(r_{+}-r\right)^{1+\bar{\alpha}} d \tilde{u} d \tilde{v},
\end{equation}
 
 However, this is not the end of the story. Similar to the Schwarzschild case, the Jacobian and its higher-order relatives will be undefined at the horizons, thus it is not straightforward to take the derivatives of the conformal factor implicitly. This is some kind of warning to the reader to avoid inferring that the modulus invoked in equation (\ref{I-Type Metric}) means that the metric is not differentiable there. At the end of the day, the Jacobian itself is ill-defined or invertable near to the horizons.  In other words, to check the differentiability of the metric in these coordinates, the metric should be fully written in terms of the CCG Kruksal coordinate first, as this will invoke generalized Lambert functions. This might be easier to handle numerically. However, no matter if these kinks cause a non-differentiablility of the metric or not near the horizons, we can always get rid of them; for instance via the use of relaxation functions.
 \vspace{1mm}

 In short, another set of coordinates $(\tilde{u},\tilde{v})$ can be introduced which will inherit the properties mentioned above and possess a relaxed (without Kinks) conformal factor at $r_{+}$ and $r_{-}$. As a consequence, the metric will be guaranteed to be $C^\infty$ if these Kinks were the only problem in these old coordinates. We list the relaxed version of the example of type-I in Appendix \ref{Appendix B}. Before we move to construct the type-II coordinates $\mathcal{GK_{II}}$, there are two features of the metric worth commenting on. First, the metric is not asymptotically flat and is different from the $\mathcal{K_{\pm}}$ coordinates where the induced metric on the submanifold $M_2=\mathcal{M}\symbol{92} SO(3)$ is asymptotically vanishing. In $\mathcal{GK_{I}}$ coordinates the induced metric on $M_2$ blows up. This is completely natural as the coordinates are compact, hence the proper distances is invariant. Second, the $\mathcal{GK_{I}}$ coordinates are dynamically casting \cite{Wei_private,wei_2023} the metric since the conformal factor includes explicit time dependence after and before the relaxation. This prevents the $r$ and $t$ from being related to $(\tilde{u},\tilde{v})$  by simple transformation similar to (\ref{radialplus}, \ref{radialminus}).
 
\subsection{Type-II Global Chart}\label{II-Type Global Chart}
While constructing $\mathcal{GK_{I}}$, a simple zero at each horizon $r_{\pm}$ was a coupled one at the other horizon $r_{\mp}$. This product of zeros had a semi-positive regular amplitude everywhere as shown in equation (\ref{I-Type Metric}) or  (\ref{I-Type Metric modified}). However, for $\mathcal{GK_{II}}$ we will have a different singularity structure that serves the same purpose: sum of two zeros at each horizon $r_{\pm}$, each coupled to a semi-positive singular amplitude at the other horizon $r_{\mp}$ that is singular at the other horizon. In principle, this class of charts should contain families of coordinates at which the coordinates themselves are extrapolation between the two Outer and Inner Kruksal coordinates. In light of this statement, the chart given in \cite{soltani_2023} plausibly belongs to that class, for more details look Appendix \ref{Appendix A}.
\vspace{1mm}

The conformal metric will have a simple pole at $r=r_{\pm}$ coupled to $M_{\pm}(r_{*},t)$, while $\beta\left(r_*, t\right)$ is effectively a residual term for completeness. $M_{\pm}(r_{*},t)$ and $\beta(r_{*},t)$ are satisfying the following constraints. As $r \rightarrow r_{\pm}$

    \begin{equation}\label{constraints of gammas and beta}
\begin{aligned}
M_{\pm}\left(r_*, t\right) &\rightarrow constant, \\
M_{\mp}\left(r_*, t\right) &\rightarrow 0, \\
\beta\left(r_*, t\right) &\rightarrow 0.
\end{aligned}
\end{equation}
Alternatively, we can restate the first constraint as: $M_{\pm}$ must have no overall pole at $r_{+}$ ($r_{-}$). Later, through this analysis, we will learn that $\beta$ will be the key to finding the global conformal charts in this procedure for the type-II coordinates. Given equation (\ref{RN tortoise coordinate}), we can rewrite this in terms of the $\mathcal{K_{\pm}}$ or the double null coordinates as follows
  \begin{equation}\label{DE}
        \begin{aligned}
             \frac{dh}{du}\frac{dk}{dv}&=
             \dfrac{exp\left(\dfrac{2 (r_{*}-r_{+})}{\alpha_{+}}\right)}{\left|r-r_{-}\right|^{\alpha}} M_{+}(r_{*},t)+\beta(r_{*},t)\\
             &+\dfrac{exp\left(-\dfrac{2 (r_{*}-r_{-})}{\alpha_{-}}\right)}{\left|r-r_{+}\right|^{\bar{\alpha}}} M_{-}(r_{*},t),
        \end{aligned}
    \end{equation}
Revisiting the condition in equation (\ref{constraints of gammas and beta}), $M_{+}$ ($M_{-}$) must have zeros at $r=r_{-}$ ($r=r_{+}$) of rank higher than $\alpha$ ($\bar{\alpha}$) respectively.
\vspace{1mm}
Searching for solutions for equation ($\ref{DE}$) could be more fruitful if we were able to find functions $M_{\pm}$ and $\beta$ with $(r_{*}\pm t)$ dependence. Accordingly, the residual term $\beta$ could be used to easily factorize the right-hand side of the equation ($\ref{DE}$) into a product of $u$- and $v$-dependent functions. The task of generating a solution to equation ($\ref{DE}$) is not trivial, but if we find $M_{\pm}(u,v)$ and $\beta(u,v)$, this will boost our progress towards achieving this task. The easiest hint we can get from the form of that equation is to try to construct $M_{+}(M_{-})$ from the $\mathcal{K_{+}}$ ($\mathcal{K_{-}}$).
Following this logic, using the trial and error method, we learn that if we define $M_{\pm}$ as  
\begin{equation}
\begin{aligned}
M_{+}&\equiv \frac{\mu_u \mu_v}{\left(1+U_{+}^{1+2 \bar{\alpha}}\right)\left(1+V_{+}^{1+2 \bar{\alpha}}\right)}\\
M_{-}&\equiv   \frac{\mu_u \mu_v}{\left(1+U_{-}^2\right)\left(1+V_{-}^2\right)},
\end{aligned}
\end{equation}
we can find $\beta$ that can do the factorization for us
\begin{equation}
\beta \equiv \frac{\mu_u \mu_v U_{+} V_{-}}{\left(1+U_{+}^{1+2 \bar{\alpha}}\right)\left(1+V_{-}^2\right)}+\frac{\mu_u \mu_v  U_{-} V_{+}}{\left(1+V_{+}^{1+2 \bar{\alpha}}\right)\left(1+U_{-}^2\right)}.
\end{equation}
This will leave us eventually with the following choices for $\dfrac{dh}{du}$ and $\dfrac{dk}{dv}$
\begin{equation}
\begin{aligned}
\frac{d h}{d u}&=\mu_u \left[\frac{U_{+}}{1+U_{+}^{1+2 \bar{\alpha}}}+\frac{U_{-}}{1+U_{-}^2}\right]\\
\frac{d k}{d v}&=\mu_v \left[\frac{V_{+}}{1+V_{+}^{1+2 \bar{\alpha}}}+\frac{V_{-}}{1+V_{-}^2}\right]\\
\end{aligned}
\end{equation}
We can also integrate these ODEs in terms of the hypergeometric and arctan functions.
\begin{equation}
\begin{gathered}
S_{\pm}(x)=\alpha_{-} \tan ^{-1}\left(e^{\frac{x}{\alpha_{-}}}\right)\\
+\bar{\alpha}_{ \pm} e^{\frac{x}{\bar{\alpha}_{ \pm}}}{ }_2 F_1\left(1,1-\bar{k}_{\pm}, 2-\bar{k}_{\pm},-e^{\bar{\kappa} x}\right)
\end{gathered}
\end{equation}
where 
\begin{equation}
\begin{gathered}
\bar{k}_{\pm}=\frac{\bar{\alpha}_{\pm}}{\bar{\alpha}_{+}+\bar{\alpha}_{-}}\\
\bar{\kappa}=\frac{1}{\bar{\alpha}_{-}}+\frac{1}{\bar{\alpha}_{+}}\\
\bar{\alpha}_{-}=\frac{\alpha_-}{2}\\
\bar{\alpha}_{+}=\alpha_+
\end{gathered}
\end{equation}
The definition for the ($\tilde{u}$,$\tilde{v}$) follows in the same way as type-I ones in (\ref{definitions of type-I in HyperGeo}). One more time, the choice we made for $\mathcal{GK_{II}}$ coordinates is naturally compact which means we can use these coordinates directly to build the Penrose diagrams 
\begin{figure}
\includegraphics[width=8cm]{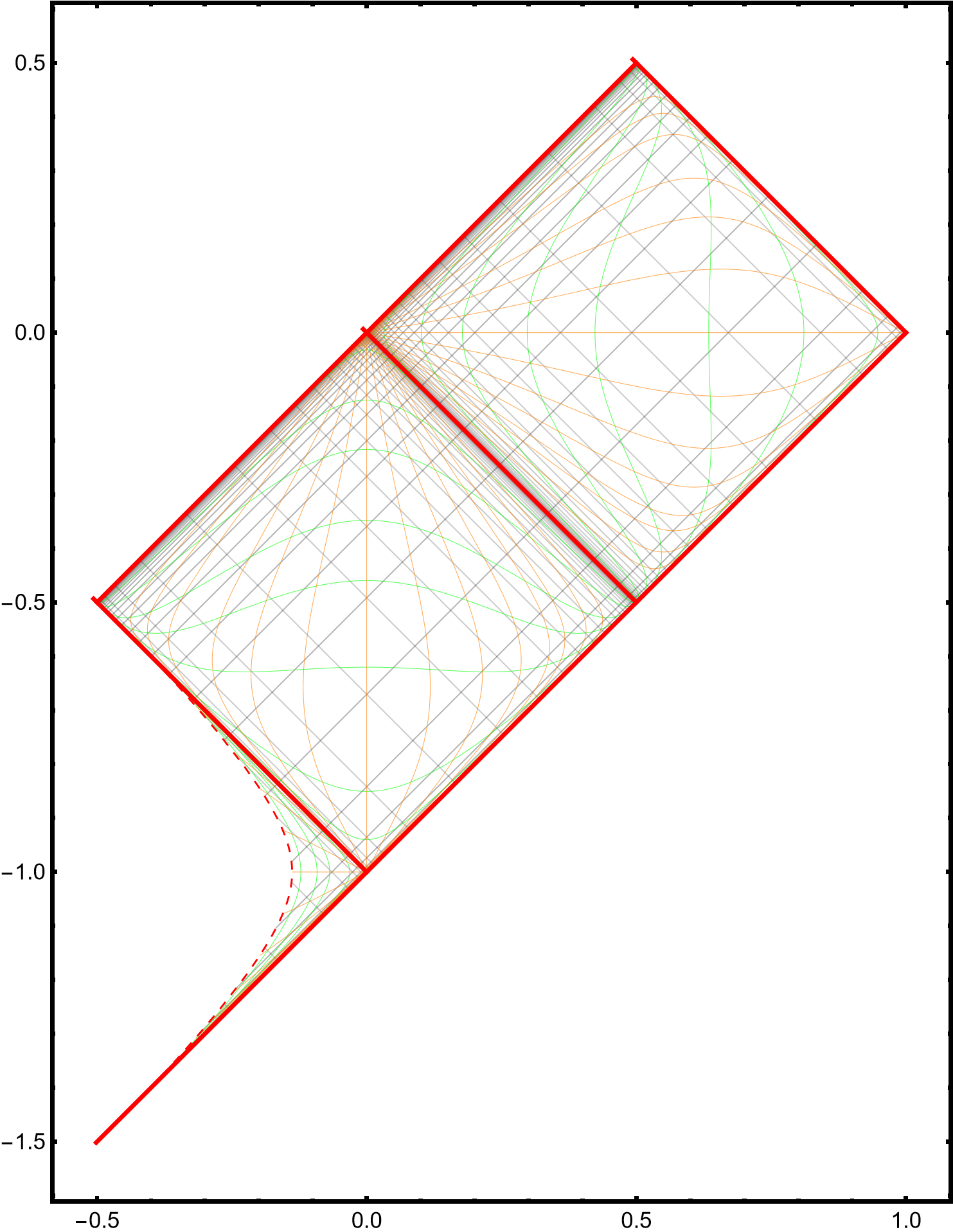}
\caption{Penrose Diagram for a single universe of Reissner Nordstrom ($M=1$, $Q=.96$) Universes using type-II coordinates. The Kruskal coordinates $(T,R)$ are plotted on the $y$-axis and $x$-axis respectively. The constant-$r$ and constant-$t$ curves are plotted in green and orange while the null-geodesics are plotted in gray. The outer $r_{+}$ and inner $r_{-}$ horizons are described by $T= R$ and $T=- R-1$ lines. While the physical singularity at $r=0$ is plotted as a red curve.
\label{PD for single Universe Type-II}}
\end{figure}. Again, our choice of the integration reference point will be the outer horizon $r_{+}$. We can now write the metric 
\begin{equation}\label{II-Type Metric}
\begin{aligned}
d S_{RN}^2
&=-\frac{1}{r^2}\left \{\ A_{+}^{-1}(r,t)+A_{-}^{-1}(r,t)+A^{-1}(r,t) \right\}^{-1} d \tilde{u} d \tilde{v}\\
\end{aligned}
\end{equation}
where $A_{\pm}(r,t)$ and $A(r,t)$ are defined as follows
\begin{equation}
\begin{aligned}
A_{+}(r,t)&\equiv exp\left(-\frac{2 r}{\alpha_{+}}\right)\left|r-r_{-}\right|^{\alpha+1}\\
&+ exp\left(\frac{2r}{\alpha_{-}}\right)\left|r-r_{+}\right|^{1+2\bar{\alpha}}\left|r-r_{-}\right|^{-1}\\
&+2 \cosh \left[t\left(\frac{1}{\alpha_{+}}+\frac{2}{\alpha_{-}}\right)\right]\\
&\times exp\left(r\left[-\frac{1}{\alpha_{+}}+\frac{2}{\alpha_{-}}\right]\right)\left|r-r_{+}\right|^{\bar{\alpha}+\frac{1}{2}}  \left|r-r_{-}\right|^\frac{\alpha}{2}\\
\end{aligned}
\end{equation}
\begin{equation}
    \begin{aligned}
 A_{-}(r,t)&\equiv exp\left(\frac{2 r}{\alpha_{-}}\right)\left|r-r_{+}\right|^{\bar{\alpha}+1}\\
&+exp\left(-\frac{2 r}{\alpha_{-}}\right)\left|r-r_{-}\right|^2\left|r-r_{+}\right|^{-\bar{\alpha}+1}\\
&+2 \cosh \left[\frac{2 t}{\alpha_{-}}\right]\left|r-r_{-}\right|\left|r-r_{+}\right|\\  
    \end{aligned}
\end{equation}
\begin{equation}
    \begin{aligned}
   A(r,t)&\equiv 2\cosh\left[\kappa t\right] \exp (-\bar{\kappa} r)\left|r-r_{+}\right|^{\frac{1+\bar{\alpha}}{2}}|r-r_{-}|^{\frac{1+\alpha}{2}}\\
&+2 \cosh\left[\bar{\kappa} t\right] exp \left(-\kappa r\right)\left|r-r_{+}\right|^{\frac{1-\bar{\alpha}}{2}}|r-r_{-}|^{\frac{3+\alpha}{2}}\\
&+2 \cosh\left[\frac{-t}{\alpha_{-}} \right] exp \left(\frac{3r}{\alpha_{-}} \right)\left|r-r_{+}\right|^{\frac{2+3\bar{\alpha}}{2}}|r-r_{-}|^{\frac{-1}{2}}\\
&+2 \cosh\left[\frac{-3t}{\alpha_{-}} \right] exp \left(\frac{r}{\alpha_{-}} \right)\left|r-r_{+}\right|^{\frac{2+\bar{\alpha}}{2}}|r-r_{-}|^{\frac{1}{2}} ,   
    \end{aligned}
\end{equation}

with the following limits
\begin{equation}
\begin{aligned}
& A_{+}\left(r \rightarrow r_{+}, t\right) \rightarrow exp\left(-\frac{2 r}{\alpha_{+}}\right)\left|r-r_{-}\right|^{\alpha+1}, \\
& A_{-}\left(r \rightarrow r_{-}, t\right) \rightarrow exp\left(\frac{2 r}{\alpha_{-}}\right)\left|r-r_{+}\right|^{\bar{\alpha}+1}, \\
& A_{\pm}\left(r \rightarrow r_{\mp}, t\right) \rightarrow \infty, \\
& A\left(r \rightarrow r_{\pm}, t\right) \rightarrow \infty.
\end{aligned}
\end{equation}
Consequently, the metric will take the following asymptotic limit
\begin{equation}
\begin{aligned}
    d S_{RN}^2(r\rightarrow r_{+})&= - \frac{e^{\frac{-2 r}{\alpha_{+}}}|r-r_{-}|^{1+\alpha}}{r^2} d \tilde{u} d \tilde{v},\\
    d S_{RN}^2(r\rightarrow r_{-})&= -  \frac{e^{\frac{2 r}{\alpha_{-}}}|r-r_{+}|^{1+\bar{\alpha}}}{r^2} d \tilde{u} d \tilde{v}.\\
\end{aligned}
\end{equation}

\begin{figure}
\includegraphics[width=8cm]{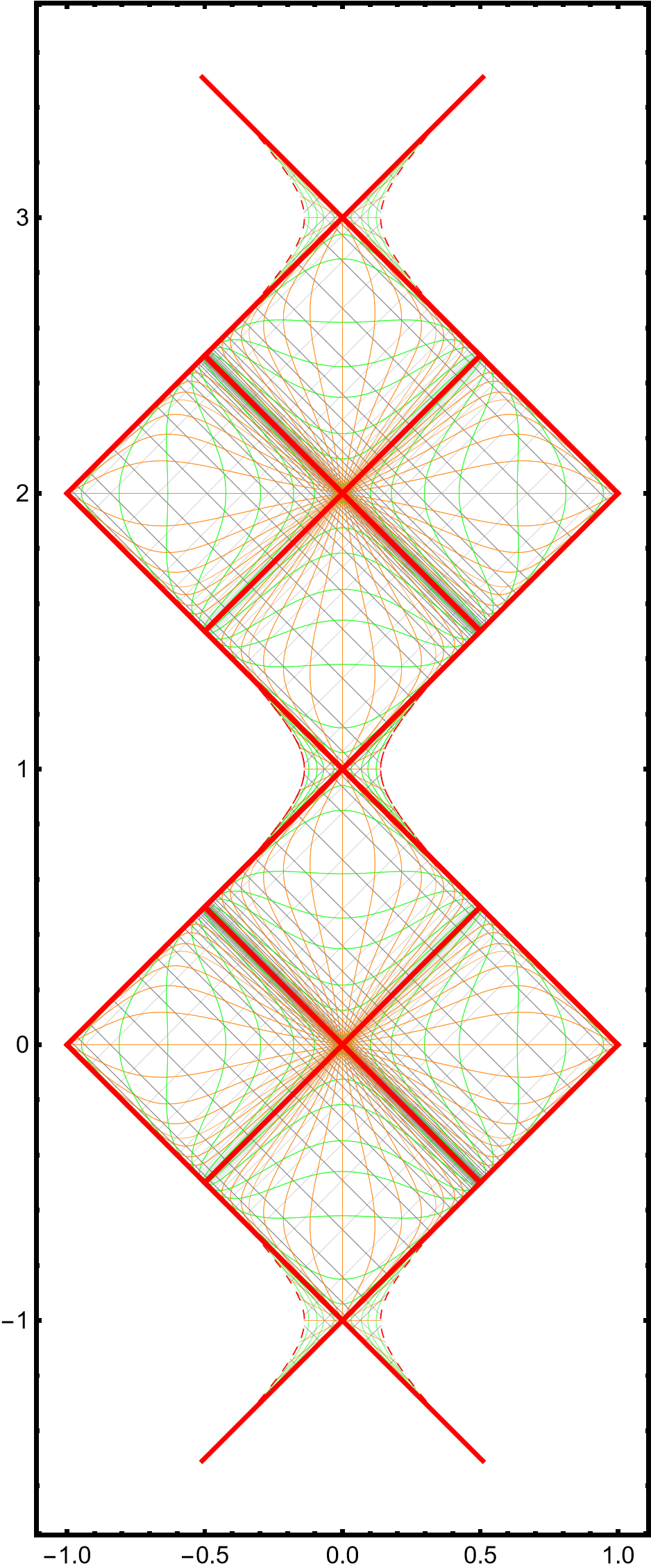}
\caption{Penrose Diagram for a maximal analytical extension of Reissner Nordstrom ($M=1$, $Q=.96$) using type-II coordinates. The Kruskal coordinates $(T,R)$ are plotted on the $y$-axis and $x$-axis respectively. The constant-$r$ and constant-$t$ curves are plotted in green and orange while the null-geodesics are plotted in gray. The outer $r_{+}$ and inner $r_{-}$ horizons are described by $T=\pm R \pm constant$ and $T=\pm R\pm1  \pm constant$ lines. While the physical singularity at $r=0$ is plotted as a dashed red curve.
\label{PD for Maxmimal Universe Type-II}}
\end{figure}.

\section{Discussion and Conclusion}\label{Discussion}

After reinterpreting the premises of the Kruskal charting of the Schwarzschild spacetime, we were able to provide a new approach to chart the Reissner–Nordström spacetime featuring two horizons. The technique itself showed to be employable in two distinctive ways, resulting in two families of charting systems: conformal global \emph{type-I} and \emph{type-II} charts. In both cases, the asymptotic form of the metric approaches the form of metric written in terms of \emph{Type-O} charts. We illustrated the success of the provided technique by constructing compact conformal global coordinates of type-I $\mathcal{GK_{I}}$ and of type-II $\mathcal{GK_{II}}$ for the RN spacetime. The price we pay for covering the whole spacetime with two horizons with only one chart is time dependence \cite{Wei_private,wei_2023}.
\vspace{2mm}

After the construction, for both type-I and type-II charts, the metric becomes $C^\infty$ if the kinks in the conformal factor near the horizons cause any ill-differentiability through the extra \emph{relaxation} step in the procedure, we give an example of how to apply this step to the example of type-I chart in appendix \ref{Appendix B}.  As expected, it is complicated to write the generalized Kruskal coordinates $(\tilde{u},\tilde{v})$ explicitly in terms of the RN coordinates $(t,r)$,  related through equation ($\ref{radialplus},\ref{radialminus}$). However, we hinted that this could be achieved by utilizing the generalized Lambert function $\mathcal{W}$ in similar manner to the use of the Lambert function $W$ in Schwarzchild case.
\vspace{2mm}

For the charts we have provided of both types, we found that the Hypergeometric functions could be employed to map the non-global null coordinate to type-I and type-II CCG charts. Finally, we demonstrated that the smoothing technique developed in \cite{soltani_2023} could be thought of as a special case of the type II family of coordinates, as could be found in Appendix \ref{Appendix A}. We believe that it is straightforward to apply this technique to spherical symmetric spacetimes with two horizons including non-strong spherical symmetric ones. Whether this procedure is applicable to glued spacetimes through thin shell approximation or to only the axial symmetric Kerr is fuzzy to authors and we believe separate and further analysis is required in order to answer such questions.

\appendix
\section{Soltanti's Smoothing Technique as Type-II}\label{Appendix A}
\begin{figure}
\includegraphics[width=8cm]{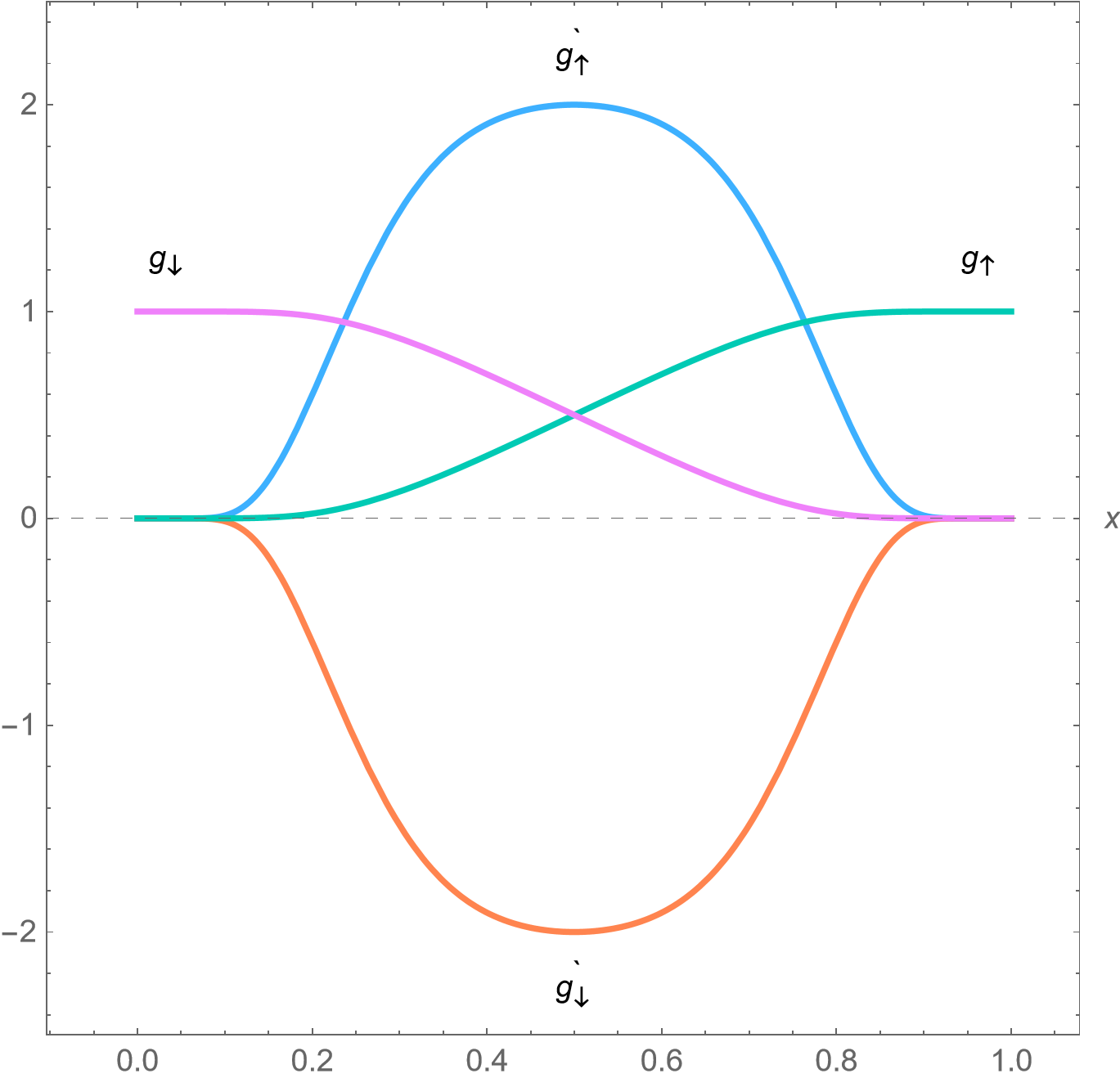}
\caption{the $g_{\uparrow}(x)$ and $g_{\downarrow}(x)$ and their derivatives of the $g_{\uparrow}^{\prime}(x)$ and $g_{\downarrow}^{\prime}(x)$.
\label{derivative of the bump functions}}
\end{figure}

Following the notation in \cite{soltani_2023} and starting from equations (31-34), the metric could be written as follows
\begin{equation}
\begin{aligned}
d S_{R N}^2&=\frac{\left(r-r_{+}\right)\left(r-r_{-}\right)}{r^2}dU dV\left\{\frac{g_{\downarrow}(U) g_{\downarrow}(V)}{k_{+}^2 U V}\right.\\
&\left.+\frac{g_{\uparrow}(U) g_{\uparrow}(V)}{k^2(U-1)(V-1)}+\frac{M(U,V)}{k_{+} k_{-}}\right\} +r^2 d \Omega^2
\end{aligned}
\end{equation}
where 
\begin{equation}
\begin{aligned}
M(U,V)=&M_1(U,V)+M_2(U,V)+M_3(U,V)\\
M_1(U,V)=&\frac{g_{\downarrow}(U) g_{\uparrow}(V)}{U(V-1)}+\frac{g_{\uparrow}(U) g_{\downarrow}(V) }{V(U-1)}\\
M_2(U,V)=& g^{\prime}_{\downarrow}(U) g^{\prime}_{\downarrow}(V) ln|U| ln|V|\\
+&g^{\prime}_{\uparrow}(U) g^{\prime}_{\uparrow}(V)ln|U-1| ln|V-1|\\
+&g^{\prime}_{\uparrow}(U) g^{\prime}_{\downarrow}(V)ln|U-1| ln|V|\\
+&g^{\prime}_{\downarrow}(U) g^{\prime}_{\uparrow}(V)ln|U| ln|V-1|\\
M_3(U,V)=&g_{\downarrow}^{\prime}(U)\left\{\frac{g_{\downarrow}(V)}{V}+\frac{g_{\uparrow}(V)}{V-1}\right\}ln|U| \\
+& g_{\uparrow}^{\prime}(U)\left\{\frac{g_{\downarrow}(V)}{V}+\frac{g_{\uparrow}(V)}{V-1}\right\}ln|U-1|\\
+& g_{\downarrow}^{\prime}(V)\left\{\frac{g_{\downarrow}(U)}{U}+\frac{g_{\uparrow}(U)}{U-1}\right\}ln|V|\\
+& g_{\uparrow}^{\prime}(V)\left\{\frac{g_{\downarrow}(U)}{U}+\frac{g_{\uparrow}(U)}{U-1}\right\}ln|V-1|\\
\end{aligned}
\end{equation}
First let us discuss the behaviour of $M$ near the horizons. As we can see from figure [\ref{derivative of the bump functions}] the $M_2$ function will be vanishing near to each horizon which are located at ${0,1}$ of the coordinates $U$($V$). Similarly we can approximate equation (31-32) and only keeping the poles near each horizon to study the behaviour of $M_1$, as $r \rightarrow r_{+}$
\begin{equation}\label{asymtotic of 31}
\begin{aligned}
&e^{k_{+} u} \simeq\frac{1}{U} \propto \frac{1}{\left|r-r_{+}\right|^{\frac{1}{2}} \left|r-r_{-}\right|^\frac{-1}{2 k_{-}}}  \\
&e^{-k_{+} v}  \simeq\frac{1}{V} \propto \frac{1}{\left|r-r_{+}\right|^{\frac{1}{2}} \left|r-r_{-}\right|^\frac{-1}{2 k_{-}}} 
\end{aligned}
\end{equation}
and as $r \rightarrow r_{-}$
\begin{equation}\label{asymtotic of 32}
\begin{aligned}
&e^{k_{-}u}  \simeq \dfrac{1}{U-1} \propto \dfrac{1}{\left|r-r_{-}\right|^{\frac{1}{2}} \left|r-r_{+}\right|^\frac{-1}{2 k_{+}}} \\
&e^{-k_{-}v} \simeq\dfrac{1}{V-1} \propto \dfrac{1}{\left|r-r_{-}\right|^{\frac{1}{2}} \left|r-r_{+}\right|^\frac{-1}{2 k_{+}}}
\end{aligned}
\end{equation}
Consequently, the product of $h(r) M_1$ is extrapolating to zero near to each horizon. Finally, a similar argument applies to $h(r) M_3$ with the extra piece of information that $g_{\downarrow(\uparrow)}^{\prime}(x)$ growth rate is exponential approach the zero near the horizons which will overwhelm the $ln |U(V)|$ blowing up there, thus this product will also extrapolate to zero near horizons. Accordingly, $h(r) M \rightarrow 0$ in agreement to the conditions we imposed on type-II coordinate $\beta$ function. 

Based on equations (\ref{asymtotic of 31}, \ref{asymtotic of 32}), it is easy to show that the terms $\dfrac{1}{U V}$ and $\dfrac{1}{(U-1) (V-1)}$ will follow the behaviour imposed on $M_{\pm}$. 

.

\section{Relaxed Version of Type-I CCG example}\label{Appendix B}
 We choose the function $\tanh{x}$ to do this job. The relaxed coordinate transformation is
\begin{equation}\label{I-Type modified}
\begin{gathered}
\frac{d h}{d u}=\frac{\mu_u}{\tanh \left[U_{-}^2\right] U_{+}^{-1}+\tanh \left[U_{+}^2\right] U_{-}^{-1}},\\
\frac{d k}{d v}=\frac{\mu_v}{\tanh \left[V_{-}^2\right] V_{+}^{-1}+\tanh \left[V_{+}^2\right] V_{-}^{-1}}.
\end{gathered}
\end{equation}
The metric now becomes
\begin{equation}\label{I-Type Metric modified}
\begin{aligned}
&d S_{RN}^2
=-\mu_u \mu_v\frac{d \tilde{u} d \tilde{v}}{r^2}\left \{Q_{-}(r,t)\left|r-r_{-}\right|^{\alpha+1} exp\left(-\frac{2 r}{\alpha_{+}}\right)\right.\\
&\left.+Q_{+}(r,t)\left|r-r_{+}\right|^{\bar{\alpha}+1} exp\left(\frac{2 r}{\alpha_{-}}\right)\right.\\
&\left.+ \tilde{Q}(r,t)exp\left(r\left[\frac{-1}{\alpha_{+}}+\frac{1}{\alpha_{-}}\right]\right) \frac{\left|r-r_{+}\right|^{\frac{\bar{\alpha}+1}{2}}}{\left|r-r_{-}\right|^{-\frac{\alpha+1}{2}}} \right \},\\
\end{aligned}
\end{equation}
where $Q_{-}$, $Q_{+}$, and $\tilde{Q}$ are defined as 
\begin{equation}
\begin{aligned}
    &Q_{-}(r,t)=\tanh \left[exp\left(\frac{2 (t-r)}{\alpha_{-}}\right)\frac{\left|r-r_{-}\right|}{\left|r-r_{+}\right|^{\bar{\alpha}}}\right]\\
    &\times\tanh \left[exp\left(\frac{-2 (t+r)}{\alpha_{-}}\right)\frac{\left|r-r_{-}\right|}{\left|r-r_{+}\right|^{\bar{\alpha}}}\right]\\
    &Q_{+}(r,t)=\tanh \left[exp\left(\frac{2 (r-t)}{\alpha_{+}}\right)\frac{\left|r-r_{+}\right|}{\left|r-r_{-}\right|^{\alpha}}\right]\\
    &\times\tanh \left[exp\left(\frac{-2 (-t+r)}{\alpha_{+}}\right)\frac{\left|r-r_{+}\right|}{\left|r-r_{-}\right|^{\alpha}}\right]\\
    &\tilde{Q}(r,t)=Q_{1}(r,t)exp\left(t\left(\frac{1}{\alpha_{+}}+\frac{1}{\alpha_{-}}\right)\right)\\
    &+Q_{2}(r,t)exp\left(-t\left(\frac{1}{\alpha_{+}}+\frac{1}{\alpha_{-}}\right)\right)\\
    &Q_{1}(r,t)=\tanh \left[exp\left(\frac{2 (t-r)}{\alpha_{-}}\right)\frac{\left|r-r_{-}\right|}{\left|r-r_{+}\right|^{\bar{\alpha}}}\right]\\
    &\times\tanh \left[exp\left(\frac{-2 (-t+r)}{\alpha_{+}}\right)\frac{\left|r-r_{+}\right|}{\left|r-r_{-}\right|^{\alpha}}\right]\\
    &Q_{2}(r,t)=\tanh \left[exp\left(\frac{-2 (t+r)}{\alpha_{-}}\right)\frac{\left|r-r_{-}\right|}{\left|r-r_{+}\right|^{\bar{\alpha}}}\right]\\
    &\times\tanh \left[exp\left(\frac{2 (r-t)}{\alpha_{+}}\right)\frac{\left|r-r_{+}\right|}{\left|r-r_{-}\right|^{\alpha}}\right]\\
    \end{aligned}
\end{equation}
This relaxed version of the conformal factor is guaranteed to be without any kinks everywhere in coordinates $(\tilde{u},\tilde{v})$. The integral $I_{2}$ defining $(\tilde{u},\tilde{v})$ is given by
\begin{equation}
I_{2}=\int\frac{d x}{\tanh \left(x^2\right) x^{q+1}+\tanh \left(x^{-2 q}\right)},
\end{equation}
The $q>1$ cases could be evaluated numerically, however, analytical methods could still be helpful in studying the relation between $\mathcal{K_{\pm}}$ and $\mathcal{GK_{I}}$ at any point. This could be achieved for example by employing series expansion, as mentioned earlier. Moreover, if we manage to invert equations (\ref{I-Type modified}) to solve explicitly for the null coordinates in terms of $\mathcal{GK_{I}}$, then we could employ the generalized Lambert function to solve for $(t,r)$ explicitly as well. Such an expansion is expected to recover equations (\ref{radialminus}) and (\ref{radialplus}) near to the horizons $r=r_{-}$ and $r=r_{+}$, respectively.

\begin{acknowledgments}

We express our sincere gratitude to Dr. Wei-Chen Lin\footnote{ArchennLin[AT]gmail.com} for his invaluable guidance and unwavering support throughout the duration of this project and also for sharing his own resources for plotting Penrose diagrams. Our appreciation also extends to Dr. Joseph Schindler\footnote{jcschindler01[AT]gmail.com}, Mahmoud Mansour\footnote{mansour[AT]iis.u-tokyo.jp.ac}, and their insightful contributions to various aspects of the analysis presented in this article. We extend special thanks to Dr. Sam Powers for providing invaluable comments on the draft, and our  appreciation to Professor William H. Kinney for sharing insightful perspectives on how we can plot the Penrose diagrams. The visual richness of this work owes much to the artistic talents of Haidi Fawzi\footnote{behance.net/haidyfawzi?}, whose illustrations greatly enhance the overall presentation. We owe Haidi Fawzi a considerable debt of thanks. Additionally, we acknowledge the partial support of Dr. D.S. by the US National Science Foundation under Grants No. PHY-2014021 and PHY-2310363.
\end{acknowledgments}
\bibliography{main}

\begin{thebibliography}{10}

\bibitem{soltani_2023}
Farshid Soltani.
\newblock Global kruskal-szekeres coordinates for reissner-nordstr\"om spacetime, 2023.
\newblock \href {https://arxiv.org/abs/2307.11026} {\path{arXiv:2307.11026}}.

\bibitem{RN_metric_2016_Jonatan_Nordebo}
Jonatan Nordebo.
\newblock the reissner nordström metric, Mar 2016.
\newblock URL: \url{https://www.semanticscholar.org/paper/The-Reissner-Nordstr%C3%B6m-metric-Nordebo/9311976210064b18e4f59223665cdb14cffb6c26}.

\bibitem{carroll_2003}
Sean~M. Carroll.
\newblock {\em Spacetime and Geometry: An Introduction to General Relativity}.
\newblock Cambridge University Press, Jan 2003.

\bibitem{griffiths_podolský_2012}
Jerry~B. Griffiths and Jiří Podolský.
\newblock {\em Exact Space-Times in Einstein's General Relativity}.
\newblock Cambridge University Press, Aug 2012.

\bibitem{matthias_blau}
Matthias Blau.
\newblock General relativity lecture notes.
\newblock URL: \url{http://www.blau.itp.unibe.ch/Lecturenotes.html}.

\bibitem{chandrasekhar_1983}
Subrahmanyan Chandrasekhar.
\newblock {\em The Mathematical Theory of Black Holes}.
\newblock Clarendon Press, Jan 1983.

\bibitem{Matt_Visser}
Matt Visser.
\newblock The kerr spacetime: A brief introduction.
\newblock {\em arXiv}, 2008.
\newblock \href {https://doi.org/10.48550/arXiv.0706.0622} {\path{doi:10.48550/arXiv.0706.0622}}.

\bibitem{teukolsky_2015}
Saul~A. Teukolsky.
\newblock The kerr metric, Jan 2015.
\newblock URL: \url{https://arxiv.org/abs/1410.2130}.

\bibitem{MTW}
Charles~W. Misner, K.~S. Thorne, and J.~A. Wheeler.
\newblock {\em {Gravitation}}.
\newblock W. H. Freeman, San Francisco, 1973.

\bibitem{Novikovphdthesis}
I.~D. Novikov.
\newblock {\em Early Stages of the Evolution of the Universe (Doctoral Dissertation)}.
\newblock PhD thesis, Sternberg Astronomical Institute, Moscow, 1963.

\bibitem{Lemaître_1997}
Abbé~Georges Lemaître.
\newblock The expanding universe.
\newblock {\em General Relativity and Gravitation}, 29(5):641–680, 1997.
\newblock \href {https://doi.org/10.1023/a:1018855621348} {\path{doi:10.1023/a:1018855621348}}.

\bibitem{Martel_Poisson_2001}
Karl Martel and Eric Poisson.
\newblock Regular coordinate systems for schwarzschild and other spherical spacetimes.
\newblock {\em American Journal of Physics}, 69(4):476–480, 2001.
\newblock \href {https://doi.org/10.1119/1.1336836} {\path{doi:10.1119/1.1336836}}.

\bibitem{Robertson_Noonan_1969}
Howard~P. Robertson and Thomas~W. Noonan.
\newblock {\em Relativity and cosmology}.
\newblock Saunders, 1969.

\bibitem{Finkelstein_1958}
David Finkelstein.
\newblock Past-future asymmetry of the gravitational field of a point particle.
\newblock {\em Physical Review}, 110(4):965–967, 1958.
\newblock \href {https://doi.org/10.1103/physrev.110.965} {\path{doi:10.1103/physrev.110.965}}.

\bibitem{kruskal_1960}
M.~D. Kruskal.
\newblock Maximal extension of schwarzschild metric.
\newblock {\em Physical Review}, 119(5):1743–1745, 1960.
\newblock \href {https://doi.org/10.1103/physrev.119.1743} {\path{doi:10.1103/physrev.119.1743}}.

\bibitem{Unruh}
W.~G. Unruh.
\newblock Global coordinates for schwarzschild black holes.
\newblock URL: \url{http://www.theory.physics.ubc.ca/530-21/bh-coords2.pdf}.

\bibitem{lemos_silva_2021}
Jos{\'{e}}~P.S. Lemos and Diogo~L.F.G. Silva.
\newblock Maximal extension of the schwarzschild metric: From painlev{\'{e}}{\textendash}gullstrand to kruskal{\textendash}szekeres.
\newblock {\em Annals of Physics}, 430:168497, jul 2021.
\newblock URL: \url{https://doi.org/10.1016%2Fj.aop.2021.168497}, \href {https://doi.org/10.1016/j.aop.2021.168497} {\path{doi:10.1016/j.aop.2021.168497}}.

\bibitem{Campanelli_Khanna_Laguna_Pullin_Ryan_2001}
Manuela Campanelli, Gaurav Khanna, Pablo Laguna, Jorge Pullin, and Michael~P Ryan.
\newblock Perturbations of the kerr spacetime in horizon-penetrating coordinates.
\newblock {\em Classical and Quantum Gravity}, 18(8):1543–1554, 2001.
\newblock \href {https://doi.org/10.1088/0264-9381/18/8/310} {\path{doi:10.1088/0264-9381/18/8/310}}.

\bibitem{Sorge_2022}
Francesco Sorge.
\newblock Kerr spacetime in lemaitre coordinates.
\newblock {\em Arxiv}, Jan 2022.
\newblock \href {https://doi.org/arXiv:2112.15441} {\path{doi:arXiv:2112.15441}}.

\bibitem{Bambi_2020}
Cosimo Bambi.
\newblock Astrophysical black holes: A review.
\newblock {\em Proceedings of Multifrequency Behaviour of High Energy Cosmic Sources - XIII — PoS(MULTIF2019)}, 2020.
\newblock \href {https://doi.org/10.22323/1.362.0028} {\path{doi:10.22323/1.362.0028}}.

\bibitem{zajaček_tursunov_2019}
Michal Zajaček and Arman Tursunov.
\newblock Electric charge of black holes: Is it really always negligible?, 2019.
\newblock \href {https://arxiv.org/abs/1904.04654} {\path{arXiv:1904.04654}}.

\bibitem{Cardoso_2016}
Vitor Cardoso, Caio~F.B. Macedo, Paolo Pani, and Valeria Ferrari.
\newblock Black holes and gravitational waves in models of minicharged dark matter.
\newblock {\em Journal of Cosmology and Astroparticle Physics}, 2016(05):054--054, may 2016.
\newblock URL: \url{https://doi.org/10.1088%2F1475-7516%2F2016%2F05%2F054}, \href {https://doi.org/10.1088/1475-7516/2016/05/054} {\path{doi:10.1088/1475-7516/2016/05/054}}.

\bibitem{Dai:2009hx}
De-Chang Dai, Katherine Freese, and Dejan Stojkovic.
\newblock {Constraints on dark matter particles charged under a hidden gauge group from primordial black holes}.
\newblock {\em JCAP}, 06:023, 2009.
\newblock \href {https://arxiv.org/abs/0904.3331} {\path{arXiv:0904.3331}}, \href {https://doi.org/10.1088/1475-7516/2009/06/023} {\path{doi:10.1088/1475-7516/2009/06/023}}.

\bibitem{hamilton_2020}
Andrew J.~S. Hamilton.
\newblock General relativity, black holes and cosmology, Jan 2020.
\newblock URL: \url{https://jila.colorado.edu/~ajsh/astr3740_17/grbook.pdf}.

\bibitem{klösch_strobl_1996}
Thomas Klösch and Thomas Strobl.
\newblock Classical and quantum gravity in 1 + 1 dimensions. ii: The universal coverings.
\newblock {\em Classical and Quantum Gravity}, 13(9):2395--2421, sep 1996.
\newblock URL: \url{https://doi.org/10.1088%2F0264-9381%2F13%2F9%2F007}, \href {https://doi.org/10.1088/0264-9381/13/9/007} {\path{doi:10.1088/0264-9381/13/9/007}}.

\bibitem{carter_1966}
B.~Carter.
\newblock {The complete analytic extension of the Reissner-Nordstr{\"o}m metric in the special case e$^{2}$ = m$^{2}$}.
\newblock {\em Physics Letters}, 21(4):423--424, jun 1966.
\newblock URL: \url{https://ui.adsabs.harvard.edu/abs/1966PhL....21..423C}, \href {https://doi.org/10.1016/0031-9163(66)90515-4} {\path{doi:10.1016/0031-9163(66)90515-4}}.

\bibitem{graves_brill_1960}
John~C. Graves and Dieter~R. Brill.
\newblock Oscillatory character of reissner-nordstr\"om metric for an ideal charged wormhole.
\newblock {\em Phys. Rev.}, 120:1507--1513, Nov 1960.
\newblock URL: \url{https://link.aps.org/doi/10.1103/PhysRev.120.1507}, \href {https://doi.org/10.1103/PhysRev.120.1507} {\path{doi:10.1103/PhysRev.120.1507}}.

\bibitem{Wei_Geodesics}
Wei-Chen Lin and Dong-han Yeom.
\newblock {Matching Radial Geodesics in Two Schwarzschild Spacetimes (e.g. Black-to-White Hole Transition) or Schwarzschild and de Sitter Spacetimes (e.g. Interior of a Non-singular Black Hole)}.
\newblock 4 2023.
\newblock \href {https://arxiv.org/abs/2304.01654} {\path{arXiv:2304.01654}}.

\bibitem{Fazzini_Rovelli_Soltani_2023}
Francesco Fazzini, Carlo Rovelli, and Farshid Soltani.
\newblock Painlevé-gullstrand coordinates discontinuity in the quantum oppenheimer-snyder model.
\newblock {\em Physical Review D}, 108(4), 2023.
\newblock \href {https://doi.org/10.1103/physrevd.108.044009} {\path{doi:10.1103/physrevd.108.044009}}.

\bibitem{lee_2002}
John~M. Lee.
\newblock {\em Introduction to Smooth Manifolds}.
\newblock Springer, Sep 2002.

\bibitem{tu_2008}
Loring~W. Tu.
\newblock {\em An Introduction to Manifolds}.
\newblock Springer, Aug 2008.

\bibitem{Schindler_2018}
J~C Schindler and A~Aguirre.
\newblock Algorithms for the explicit computation of penrose diagrams.
\newblock {\em Classical and Quantum Gravity}, 35(10):105019, apr 2018.
\newblock URL: \url{https://doi.org/10.1088%2F1361-6382%2Faabce2}, \href {https://doi.org/10.1088/1361-6382/aabce2} {\path{doi:10.1088/1361-6382/aabce2}}.

\bibitem{Schindler_2020}
Joseph~C. Schindler, Anthony Aguirre, and Amita Kuttner.
\newblock Understanding black hole evaporation using explicitly computed penrose diagrams.
\newblock {\em Physical Review D}, 101(2), jan 2020.
\newblock URL: \url{https://doi.org/10.1103%2Fphysrevd.101.024010}, \href {https://doi.org/10.1103/physrevd.101.024010} {\path{doi:10.1103/physrevd.101.024010}}.

\bibitem{BHevaporation}
J.~C. Schindler.
\newblock {\em The Structure of Evaporating Black Holes (Doctoral Dissertation)}.
\newblock PhD thesis, UNIVERSITY OF CALIFORNIA, SANTA CRUZ, 2019.

\bibitem{doran_lobo_crawford_2008}
Rosa Doran, Francisco S.~N. Lobo, and Paulo Crawford.
\newblock Interior of a schwarzschild black hole revisited.
\newblock {\em Foundations of Physics}, 38(2):160--187, dec 2007.
\newblock URL: \url{https://doi.org/10.1007%2Fs10701-007-9197-6}, \href {https://doi.org/10.1007/s10701-007-9197-6} {\path{doi:10.1007/s10701-007-9197-6}}.

\bibitem{KS_for_f(R)}
A~Romadani and M~F Rosyid.
\newblock Kruskal-szekeres coordinates of spherically symmetric solutions in theories of gravity.
\newblock {\em Journal of Physics: Conference Series}, 1816(1):012030, feb 2021.
\newblock URL: \url{https://dx.doi.org/10.1088/1742-6596/1816/1/012030}, \href {https://doi.org/10.1088/1742-6596/1816/1/012030} {\path{doi:10.1088/1742-6596/1816/1/012030}}.

\bibitem{frolov_novikov_1998}
Valeri~P. Frolov and Igor~D. Novikov.
\newblock {\em Black hole physics : basic concepts and new developments}, volume~96.
\newblock Springer Netherlands, Jan 1998.
\newblock \href {https://doi.org/10.1007/978-94-011-5139-9} {\path{doi:10.1007/978-94-011-5139-9}}.

\bibitem{Mathmatica_notebook}
Fawzi. A.
\newblock {Generalized Kruskal Coordinate Notebook}, 2023.
\newblock URL: \url{to be updated}.

\bibitem{Wald}
Robert~M. Wald.
\newblock {\em {General Relativity}}.
\newblock Chicago Univ. Pr., Chicago, USA, 1984.
\newblock \href {https://doi.org/10.7208/chicago/9780226870373.001.0001} {\path{doi:10.7208/chicago/9780226870373.001.0001}}.

\bibitem{corless_gonnet_hare_jeffrey_knuth_1996}
Robert~M. Corless, Gaston~H. Gonnet, D.~E.~G. Hare, David~J. Jeffrey, and Donald~E. Knuth.
\newblock On the lambert w function.
\newblock {\em Advances in Computational Mathematics}, 5(1), Dec 1996.
\newblock \href {https://doi.org/10.1007/BF02124750} {\path{doi:10.1007/BF02124750}}.

\bibitem{MG2010}
Thomas Mueller and Frank Grave.
\newblock Catalogue of spacetimes, 2010.
\newblock \href {https://arxiv.org/abs/0904.4184} {\path{arXiv:0904.4184}}.

\bibitem{Mezo_Baricz_2017}
Istvan Mezo and Arpad Baricz.
\newblock On the generalization of the lambert $w$ function.
\newblock {\em Transactions of the American Mathematical Society}, 369(11):7917–7934, 2017.
\newblock \href {https://doi.org/10.1090/tran/6911} {\path{doi:10.1090/tran/6911}}.

\bibitem{Mezo_Corcino_Corcino_2020}
István Mező, Cristina~B Corcino, and Roberto~B Corcino.
\newblock Resolution of the plane-symmetric einstein-maxwell fields with a generalization of the lambert w function.
\newblock {\em Journal of Physics Communications}, 4(8):085008, 2020.
\newblock \href {https://doi.org/10.1088/2399-6528/abab40} {\path{doi:10.1088/2399-6528/abab40}}.

\bibitem{Mezo_Keady_2016}
István Mező and Grant Keady.
\newblock Some physical applications of generalized lambert functions.
\newblock {\em European Journal of Physics}, 37(6):065802, 2016.
\newblock \href {https://doi.org/10.1088/0143-0807/37/6/065802} {\path{doi:10.1088/0143-0807/37/6/065802}}.

\bibitem{Scott_Mann_Martinez_II_2006}
Tony~C. Scott, Robert Mann, and Roberto~E. Martinez~II.
\newblock General relativity and quantum mechanics: Towards a generalization of the lambert w function a generalization of the lambert w function.
\newblock {\em Applicable Algebra in Engineering, Communication and Computing}, 17(1):41–47, 2006.
\newblock \href {https://doi.org/10.1007/s00200-006-0196-1} {\path{doi:10.1007/s00200-006-0196-1}}.

\bibitem{Wei_private}
Wei-Chen Lin.
\newblock private communication.

\bibitem{wei_2023}
Wei-Chen Lin, Dejan Stojkovic, and Dong han Yeom.
\newblock Trouble with the penrose diagram in spacetimes connected via a spacelike thin shell, 2023.
\newblock \href {https://arxiv.org/abs/2302.04923} {\path{arXiv:2302.04923}}.

\bibitem{ronveaux_2007}
A.~Ronveaux.
\newblock {\em Heun's differential equations}.
\newblock Oxford University press, 2007.

\bibitem{hypergeometric1}
Hypergeometric function.
\newblock URL: \url{https://mathworld.wolfram.com/HypergeometricFunction.html}.

\bibitem{hypergeometric2}
Chapter 15 hypergeometric function.
\newblock URL: \url{https://dlmf.nist.gov/15}.

\end{thebibliography}
\end{document}